\documentclass[lettersize,journal]{IEEEtran}
\usepackage{array}
\usepackage[caption=false,font=normalsize,labelfont=sf,textfont=sf]{subfig}
\usepackage{textcomp}
\usepackage{stfloats}
\usepackage{url}
\usepackage{bm}
\usepackage{amsfonts}
\usepackage{cite}
\usepackage{verbatim}

\usepackage{graphicx}
\usepackage[export]{adjustbox} 
\usepackage{graphicx} 
\def\BibTeX{{\rm B\kern-.05em{\sc i\kern-.025em b}\kern-.08em
    T\kern-.1667em\lower.7ex\hbox{E}\kern-.125emX}}
\usepackage{balance}
\usepackage{multirow}
\usepackage{enumerate}
\usepackage{amsmath}
\usepackage{graphicx}
\usepackage{float}
\usepackage{subfloat}
\usepackage{}
\usepackage{hyperref}
\usepackage{adjustbox}
\usepackage[skip=2pt]{caption} 
\usepackage{enumitem}
\usepackage{booktabs}
\usepackage{fancyhdr} 
\usepackage{xcolor}
\usepackage[linesnumbered,ruled,vlined]{algorithm2e}
\usepackage{algorithmic}
\usepackage{caption}
\fancypagestyle{myfooter}{
  \fancyfoot[C]{\href{链接地址}{超链接内容}} 
}

\begin{document}

\title{Dual-domain Collaborative Denoising for Social Recommendation}

\author{Wenjie Chen, Yi Zhang, Honghao Li, Lei Sang, Yiwen Zhang* 
\IEEEcompsocitemizethanks{\IEEEcompsocthanksitem Wenjie Chen, Yi Zhang, Honghao Li, Lei Sang and Yiwen Zhang, are with School of Computer Science and Technology, Anhui University 230601,
Hefei, Anhui, P.R. China.   E-mail: wenjie@stu.ahu.edu.cn, zhangyi.ahu@gamil.com, salmon1802li@gmail.com, sanglei@ahu.edu.cn, zhangyiwen@ahu.edu.cn.}
\vspace{-1cm}
\thanks{*Corresponding author.}
}

\markboth{IEEE Transactions on Computational Social Systems}%
{Shell \MakeLowercase{\textit{et al.}}: Dual-domain Collaborative Denoising for Social Recommendation}

\maketitle

\begin{abstract}
Social recommendation leverages social network to complement user-item interaction data for recommendation task, aiming to mitigate the data sparsity issue in recommender systems. The information propagation mechanism of Graph Neural Networks (GNNs) aligns well with the process of social influence diffusion in social network, thereby can theoretically boost the performance of recommendation. However, existing social recommendation methods encounter the following challenge: both social network and interaction data contain substaintial noise, and the propagation of such noise through GNNs not only fails to enhance recommendation performance but may also interfere with the model's normal training.  However, despite the importance of denoising for social network and interaction data, only a limited number of studies have considered the denoising for social network and  all of them overlook that for interaction data, hindering the denoising effect and recommendation performance. Based on this, we propose a novel model called Dual-domain Collaborative Denoising for Social Recommendation ($\textbf{DCDSR}$). DCDSR comprises two primary modules: the structure-level  collaborative denoising module and the embedding-space collaborative denoising module. In the structure-level collaborative denoising module, information from  interaction domain is first employed to guide social network denoising. Subsequently, the denoised social network is used to supervise the denoising for interaction data.  The embedding-space collaborative denoising module devotes to resisting the noise cross-domain diffusion problem through contrastive learning with dual-domain embedding collaborative perturbation. Additionally, a novel contrastive learning strategy, named Anchor-InfoNCE, is introduced to better harness the denoising capability of contrastive learning. The model is jointly trained under a recommendation task and a contrastive learning task. Evaluating our model on three real-world datasets verifies that DCDSR has a considerable denoising effect, thus outperforms the state-of-the-art social recommendation methods.
\end{abstract}

\begin{IEEEkeywords}
Collaborative Denoising, Social Recommendation, Graph Neural Networks, Graph Contrastive Learning.
\end{IEEEkeywords}

\section{Introduction}
Social recommendation \cite{trustRS, socialrec} has become a significant part of recommender systems due to the rapid development of social platforms, where social network \cite{social_network1}, \cite{social_network2} plays an auxiliary role in recommedation task except user-item interaction data. Following the social homophily theory \cite{social_homophily} that users with connections in social network tend to interact with similar items, it effectively alleviates the problem of data sparsity \cite{data_sparsity1, user_cf_cold_start} in recommender systems.

\begin{figure*}[t]
    \begin{minipage}[t]{0.5\linewidth}
        \centering
        \includegraphics[width=\textwidth,height=5cm]{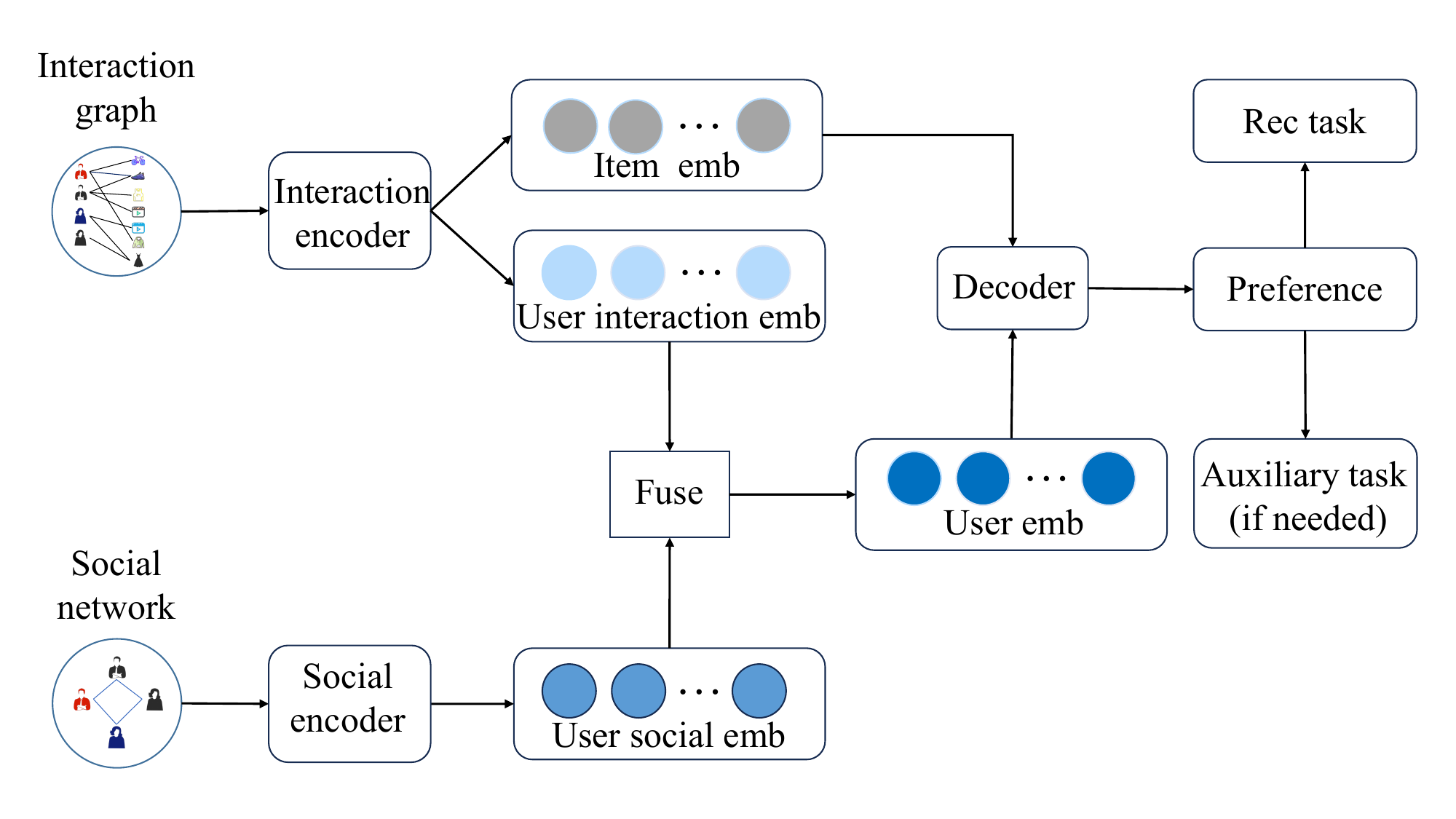}
        \caption*{(a) Graph-based social recommendation paradigm}
        \label{gsr}
    \end{minipage}%
    \begin{minipage}[t]{0.5\linewidth}
        \centering
        \includegraphics[width=\textwidth,height=5cm]{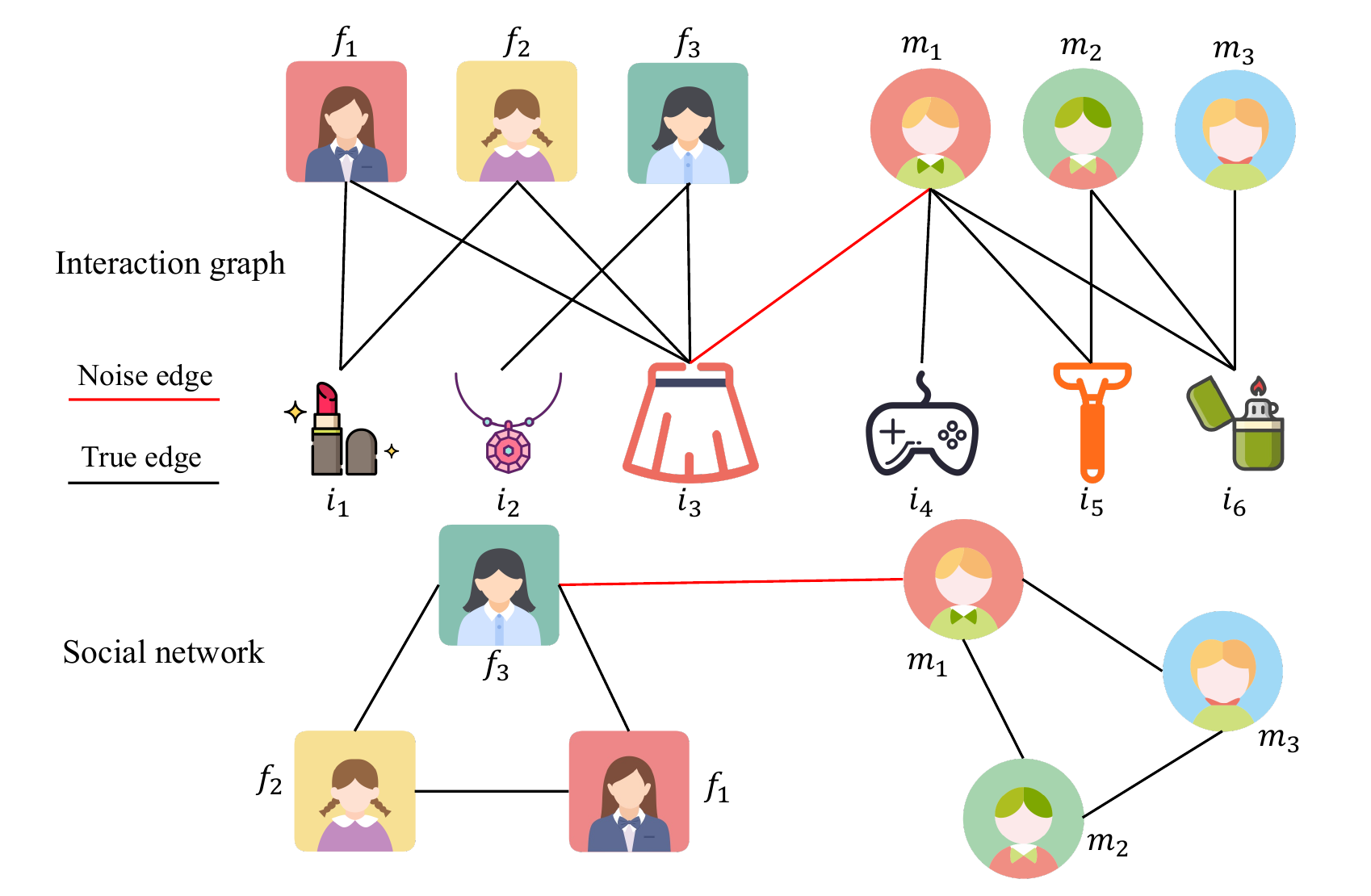}
        \caption*{(b) Motivation example}
        \label{motivation_example}
    \end{minipage}
    \begin{minipage}[t]{0.5\linewidth}
        \centering
        \includegraphics[width=\textwidth]{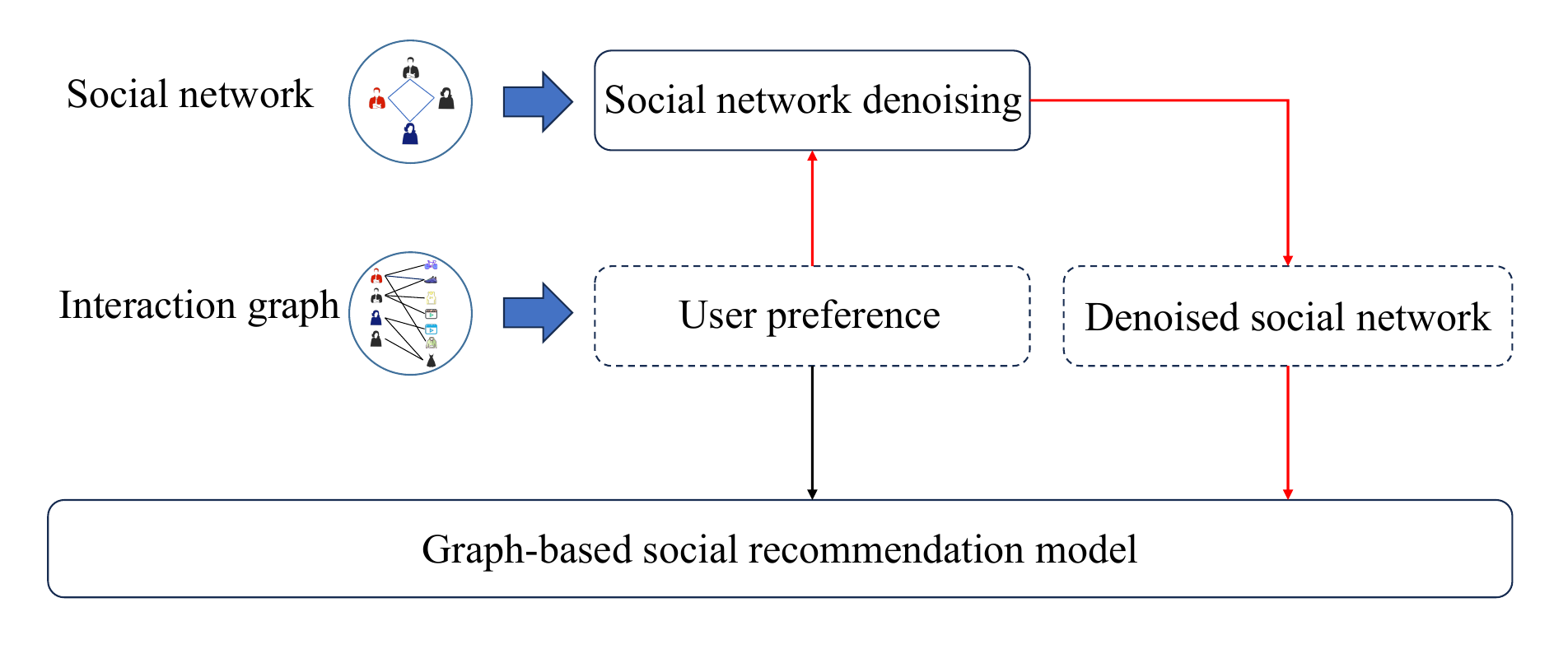}
        \caption*{(c) Preference-guided social network denoising paradigm}
        \label{normal_denoising}
    \end{minipage}%
    \begin{minipage}[t]{0.5\linewidth}
        \centering
        \includegraphics[width=\textwidth]{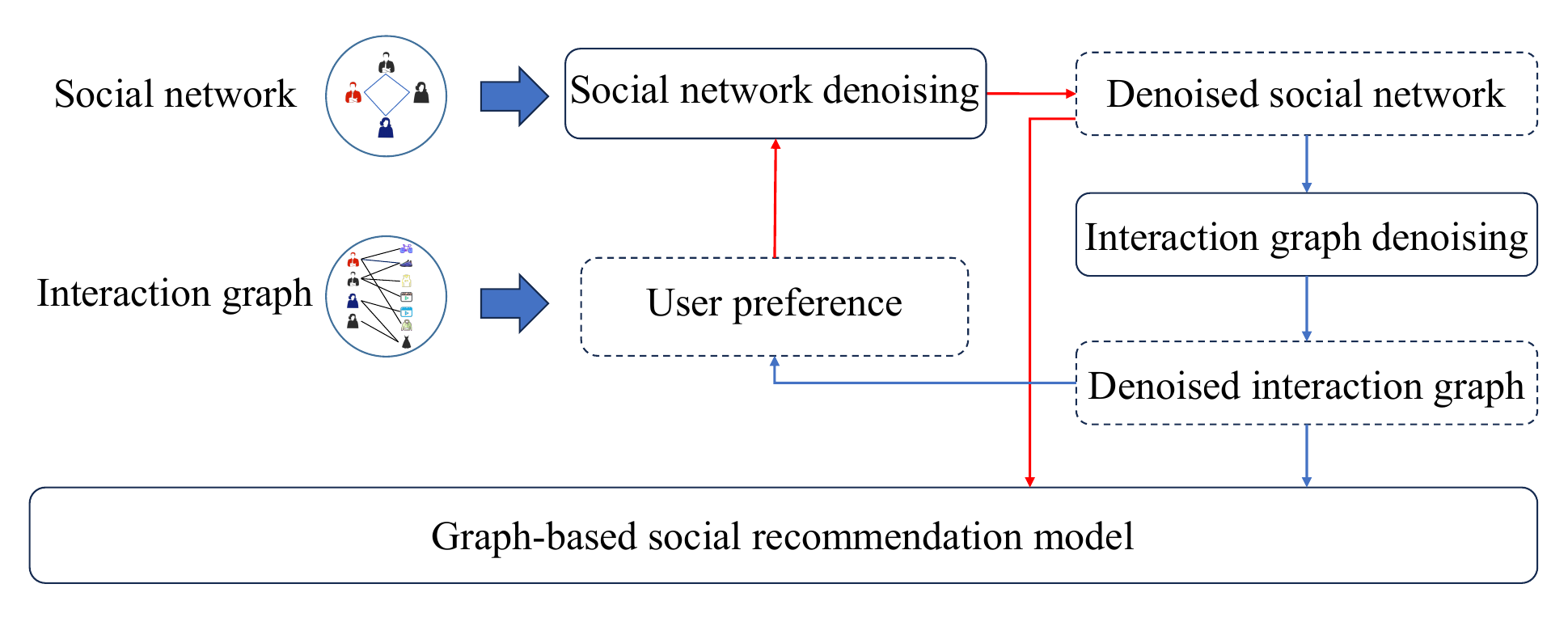}
        \caption*{(d) Dual-domain collaborative denoising paradigm}
        \label{xietongjiangzao}
    \end{minipage}
    \caption{Mainstream social recommendation paradigms for certain scenario and the motivation example for our work.}
    \label{MOTIVATION}
\end{figure*}

With the emergence of Graph Neural Networks (GNNs), Graph-based Social Recommendation (GSR) has become a mainstream paradigm (formalized as Fig. \ref{MOTIVATION}(a)) for social recommendation. This is attributed to the message propagation mechanism of GNNs, which closely aligns with the social influence process \cite{social_influence}, leading to the proposal of a series of GSR models \cite{diffnet, diffnet++, kd_gsr, social_group, DHGCN, SCGREC, social_jieou}. Following the success of Graph Contrastive Learning (GCL) across various domains, GSR leverages GCL to enable social network to provide supervised signals for interaction data \cite{mhcn,sept,dcrec}. However, a critical issue has been overlooked in these works, i.e., the presence of noise in social recommendation, encompassing both social network and interaction data.

Firstly, utilizing social network in recommender systems can often introduce amount of noise. Specifically, due to the challenges such as the difficulty of data collection, the social relations in observable social network are implicit, i.e., represented by 0/1 to indicate whether social users trust each other. However, the trust propensity theory \cite{trust-propensity} suggests that the establishment of social relations does not necessarily imply that social neighbors share similar preferences. For example, diverse students may form social relations because they are in the same class, yet their preferences can be vastly different. The analysis experiments conducted by DESIGN \cite{design} have demonstrated that such social relations account for the vast majority in the social network. Therefore, directly applying the social network to recommendation task is unwise.

Meanwhile, the noise issue is also present in user-item interaction data. Since user preference information is difficult to obtain directly, most user-item interaction data is implicit \cite{2012bpr}, i.e., in the form of implicit feedback such as clicking and browsing. However, numerous studies have highlighted the substantial discrepancy between implicit feedback and users' actual interests \cite{interaction_denoising, zhang2023revisiting, wang2021robust}, indicating that the available implicit feedback contains a significant amount of noise that is not relevant to the recommendation task. For example, if a user unintentionally clicks on a certain item page, or watches a certain type of video that he or she would not like due to his/her friend's sharing, this creates noise to implicit feedback.

Despite some efforts such as DSL \cite{dsl} and GDMSR \cite{gdmsr} have adopted preference-guided approach to denoise social network to improve the robustness of recommendation, they still suffer from a fatal flaw. That is they only focus on the denoising for social network while overlooking the noise issue in interaction data. In fact, within the preference-guided social network denoising paradigm showed in Fig. \ref{MOTIVATION}(c), the denoising effect would be vulnerable to the noisy interaction data. As illustrated in Fig. \ref{MOTIVATION}(b), there is a three-female user group ($f_1, f_2, f_3$) and a three-male user group ($m_1, m_2, m_3$), where the female user group  prefers dressing items, and the male group prefers electronic items. However, user $m_1$ passively clicks information about a dress due to a share from his social neighbor user $f_3$. This interaction slightly shifts user $m_1$'s preference towards the dressing category, causing his preference to become ambiguous. In this case, the preference-guided social network denoising paradigm  would incorrectly ignores the need to intervene with the relation $<f_3, m_1>$, which impairs the denoising effect of the social network. This issue can be described as noise cross-domain diffusion. Moreover, noisy interaction data would directly lead to suboptimal recommendation performance. 

To address above issues, we propose the model \textbf{D}ual-domain \textbf{C}ollaborative \textbf{D}enoising for \textbf{S}ocial \textbf{R}ecommendation (\textbf{DCDSR}). DCDSR aims to comprehensively address the noise issue to improve the model robustness with a dual-domain collaborative denoising paradigm which is abstracted as Fig. \ref{MOTIVATION}(d). Concretely, the \textbf{structure-level collaborative denoising module} utilizes user's preference  to perform denoising operations on the original social network, then utilizes the denoised social network in conjunction with knowledge of interaction domain to denoise the interaction graph. The denoised interaction graph can not only promote further denoising for the original social network in a cyclical manner but also positively facilitate the recommendation task. Subsequently, the \textbf{embedding-space collaborative denoising module} performs denoising for user and item embeddings based on GCL with dual-domain embedding collaborative perturbation to allievate the noise cross-domain diffusion issue existing in structure-level denoising process. Furthermore, we propose a novel contrastive learning loss \textbf{Anchor-InfoNCE} (\textbf{AC-InfoNCE}), where the embedding of user/item is set as an anchor, and the positive samples are respectively induced to align with the anchor, while different users/items are induced to converge to uniform distributions in the embedding space. This circumvents the effect of the uncertainty of data augmentation and sample distribution on GCL while expanding the number of positive samples. Above two denoising modules complement each other to cut down the noise present in the original social network and interaction graph, and finally learn high-quality embeddings for recommendation task.

Our contributions can be summarized as follows:

$\bullet$ We propose a model called Dual-domain Collaborative Denoising for Social Recommendation (DCDSR). To the best of our knowledge, DCDSR is the first work to address the noise issue present both in social domain and interaction domain with a novel dual-domain collaborative denoising paradigm.

$\bullet$  We design a structure-level collaborative denoising module which reduces the structural noise  by enabling the interaction graph and social network to mutually guide each other's cyclic reconstruction.

$\bullet$ We design a embedding-space collaborative denoising module to allievate the noise diffusion across interaction and social domains through contrastive learning with dual-domain embedding collaborative perturbation. In addition, a robust contrastive loss function AC-InfoNCE is designed to enhance the contrastive learning for embedding denoising.

$\bullet$ Extensive experiments conducted on three real-world datasets demonstrates that DCDSR is superior to those state-of-the-art GSR models and denoising GSR models. Futhermore, DCDSR is more noise-resistant when faced with varying degrees of external noise interaction attacks.

\section{PRELIMINARY AND RELATED WORKS}
In this section, we briefly review Graph-based Social Recommendation (GSR) and Graph Contrastive Learning (GCL), which are the backbones of DCDSR. In addition, for ease of reading and understanding, we list some notations with explanations in Table \ref{symbol table}.

\begin{table}[h]
\caption{Notations and explanations.}
\resizebox{\linewidth}{!}{
\begin{tabular}{c|l}
\hline \textbf{Notation} & \multicolumn{1}{c}{ \textbf{Explanation} } \\
\hline \hline $\mathcal{U}, \mathcal{I}$ & The sets of users and items. \\
$M, N$ & The numbers of users and items. \\
$\mathcal{G}_R$, $\mathcal{G}_S$  & The user-item interaction graph and social network. \\
$\mathcal{N}_u, \mathcal{S}_u$ & The first-order neighbors of user $u$ in $\mathcal{G}_R$ and $\mathcal{G}_S$. \\
$\mathbf{e}_u, \mathbf{e}_i$ & The learnable embedding of user $u$ and item $i$. \\
$\mathbf{h}_u^r, \mathbf{h}_i^r$ & The preference embedding of user $u$ and item $i$. \\
$\mathbf{h}_u^{\text{enhanced}}$ & The social-enhanced embedding of user $u$. \\
$\mathbf{p}_u^r, \mathbf{p}_u^s$ & The interaction embedding and social embedding of user $u$. \\
$\mathbf{p}_i^r$ & The interaction embedding of item $i$. \\
$PC$, $IC$ & The preference consistency and interaction compatibility. \\
$\beta_s, \beta_r$ & The denoising threshold for social network and interaction graph. \\
$\lambda_1, \lambda_2, \lambda_3$ & The weights of three CL losses. \\
\hline
\end{tabular}}
\label{symbol table}
\vspace{-1em}
\end{table}

\subsection{Graph-based Social Recommendation}
GSR is the application of GNNs based on user-item interaction graph and social network to model user and item representations for recommendation tasks. GNNs \cite{GCN, graphembedding} open the information transfer channel between graph nodes, enabling nodes to aggregate information from their first-order neighbor or higher-order neighbor nodes, thus enriching the features of nodes. In recent years, GNNs have developed rapidly and a series of graph neural network instances have been proposed, such as GCN \cite{GCN}, GAT \cite{GAT}, etc., and utilized in downstream tasks, such as node classification \cite{Nodeclass}, recommender systems \cite{NGCF}, traffic prediction \cite{traffic} and relation extraction \cite{relation_extraction}. In recommender systems, some GCNs suitable for collaborative filtering task have been proposed, e.g., NGCF \cite{NGCF}, LightGCN \cite{lightgcn}, and NIE-GCN \cite{zhang2024nie}. 

 

In GSR, user-user social network is incorporated alongside user-item interaction graph to enhance recommendation task, the logic basis of which is that a user's item preferences are simultaneously influenced by their historical item interactions and social relations \cite{trustRS, socialrec}. Thus the mainstream paradiam of GSR \cite{diffnet, diffnet++, DISGCN, sociallgn, DANSER, mhcn } separately models the user interaction embeddings and social embeddings by GNN and then explicitly or implicitly fuses these embeddings to inference user's preference for recommendation task. For example, DiffNet \cite{diffnet} organically fuses the user embeddings with those of its neighboring users via GNN and then sums them with the average embeddings of the item's neighbors to obtain the final user embedding representation. DISGCN \cite{DISGCN} decouples the user/item representations by using GNNs to learn the social homophily embeddings of users and the social influence embeddings of items on social network and interaction graph, respectively, and finally fusing these two embeddings. SocialLGN \cite{sociallgn} designs a lightweight graph convolutional network-based representation propagation mechanism for both interaction graph and social network.

Although GSR utilizes GNNs to effectively capture user-user social signals and user-item collaboration signals, its performance is still limited by noise \cite{design} issue. Therefore, some research has begun to focus on the noise issue in GSR. DSL \cite{dsl} achieves implicit denoising of social networks by adaptively aligning user similarities in both interaction and social domains. GDMSR \cite{gdmsr} models the confidence of social relations and periodically reconstructs the social network to perform denoising. These efforts are based on preference-guided social network denoising paradigm \cite{gdmsr}, which has improved the robustness of the models to some extent. However, these works have not taken into account the noise in the interaction data, thus limiting the effectiveness of social network denoising and the performance of recommendation.
\subsection{Graph Contrastive Learning for Recommendation} 
In recent years, contrastive learning has been successful in many fields (CV \cite{simclr}, NLP \cite{gao2021simcse}), and it effectively alleviates the problem of supervised signal sparsity especially on graph data, thus Graph Contrastive Learning (GCL) \cite{sgl} is increasingly being applied to graph-based recommedation including GSR. The core idea of GCL is to employ graph-augmentation (GA) or feature-augmentation (FA) and then minimize the loss such as InfoNCE \cite{infonce} (illustrated later) to achieve the mutual information maximization \cite{MIM} between the augmented samples to offer extra supervised signals. 

In graph-based recommender systems, SGL \cite{sgl} firstly utilizes GCL, which obtains two different views by randomly dropping edges/nodes on user-item interaction graph, and uses user/item embeddings in these two views for contrastive learning. SEPT \cite{sept} jointly utilizes interaction graph and social network for data augmentation, and constructs the friend view and the sharing view for contrastive learning. DcRec \cite{dcrec} introduces a novel disentangled contrastive learning framework for social recommendation that leverages contrastive learning to transfer knowledge from the social domain to the collaborative domain. SimGCL \cite{simgcl} improves the data augmentation method of SGL by directly randomly perturbing the embeddings of users/items to generate positive samples for contrastive learning, and states that the key role is played by the contrastive learning loss function InfoNCE. ReACL \cite{ReACL} harnesses multiple views of user social relations and item commonality relations to form contrastive pairs and extract relationship-guided self-supervised signals. 

GCL has significantly improved the performance of recommendation task, including GSR. However, the data augmentation methods used in GCL, have low interpretability and are uncontrollable, which can lead to the introduction of noise. Moreover, there may exist a gradient bias issue with InfoNCE during training process, which can impair the robustness of contrastive learning. Therefore, we can make further improvements to the data augmentation method and loss function to optimize the GCL paradigm and design a GCL framework that is more suitable for social recommendation scenarios, so as to better play the role of GCL in mitigating data sparsity and denoising. 

\section{DUAL-domain COLLABORATIVE DENOISING FOR SOCIAL RECOMMENDATION}
\subsection{Problem Formulation}
DCDSR aims to perform top-K item recommendations for the user set $\mathcal{U}=\left\{u_1, u_2, \ldots, u_M\right\}$ from the item set $\mathcal{I}=\left\{i_1, i_2, \ldots, i_N\right\}$. To achieve this task, we formalize the input user-item  interaction data and social trust data into a user-item interaction graph  $\mathcal{G}_R$ and a social network $\mathcal{G}_S$.

\subsection{Model Overview}
\begin{figure*}[t]
    \centering
    \includegraphics[width=1\linewidth]{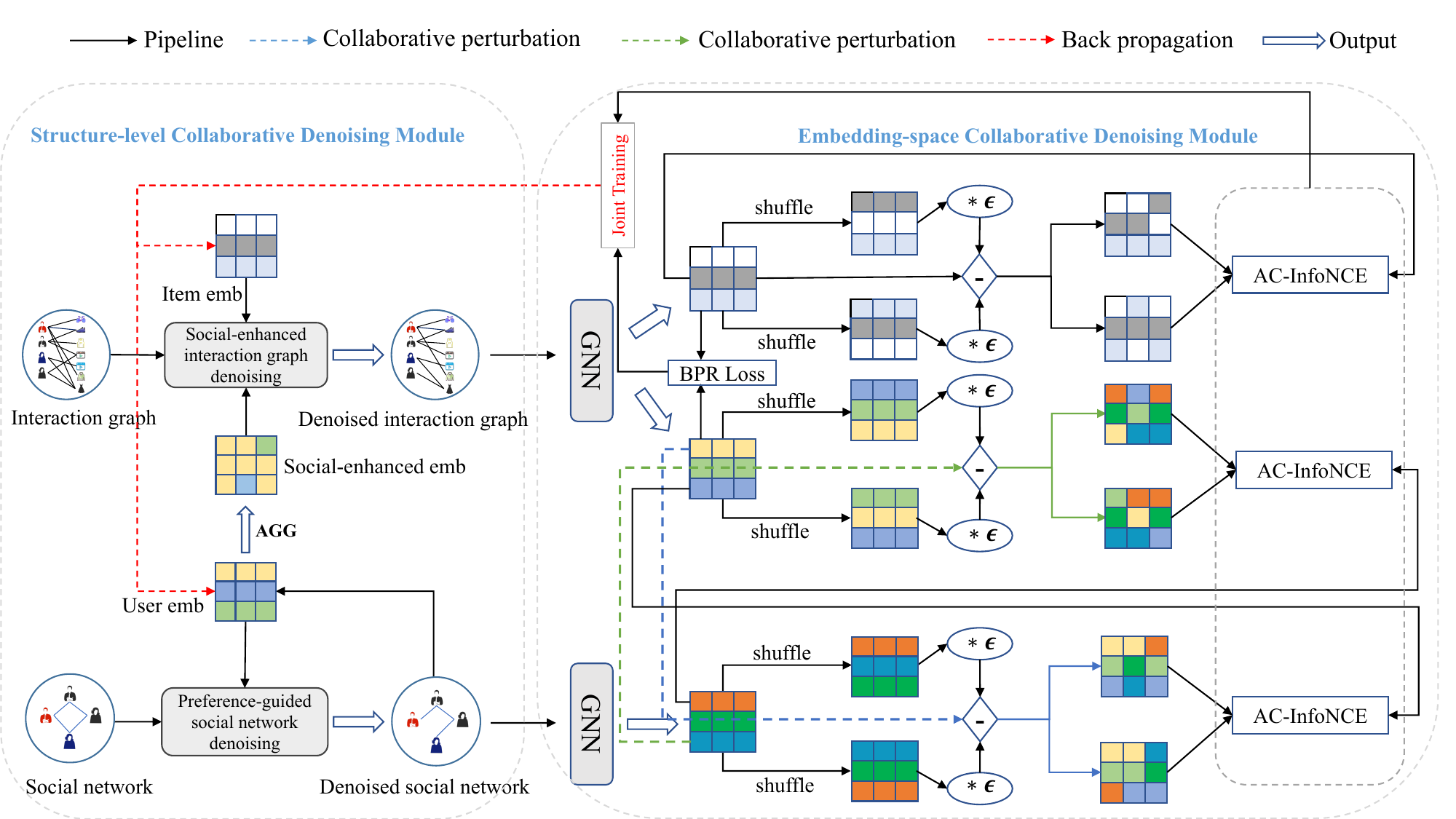}
    \caption{Overview of DCDSR.}
    \label{fig:model_fig}
\end{figure*}

As shown in Fig. \ref{fig:model_fig}, DCDSR consists of two main modules, i.e., structure-level collaborative denoising module and embedding-space collaborative denoising module. The structural-level collaborative denoising module is used to perform denoising on original social network and interaction graph to minimize the impact of structural noise on GNN learning user/item embeddings. Subsequently, the embedding-space collaborative denoising module contributes to alleviate the noise diffusion across interaction domain and social domain exisiting in structure-level collaborative denoising module. Eventually the model is trained jointly with the BPR \cite{2012bpr} loss of recommendation task and AC-InfoNCE losses of contrastive learning task.


\subsection{Structure-level Collaborative Denoising}
The goal of structure-level collaborative denoising is to evaluate the noise in the original social network $\mathcal{G}_S$ and interaction graph $\mathcal{G}_R$ and delete those noisy edges in the network/graph to obtain cleaner social network $\mathcal{G}_S'$ and interaction graph $\mathcal{G}_R'$.

\subsubsection{Preference-guided social network denoising}
In social recommendation, social network is used to assist the recommendation task which lies in there are lots of user relations and can provide extra supervised signals because social-related users tend to have consistent preferences according to the social homophily theory \cite{social_homophily}. However, many studies have shown that there are lots of social relations with dissimilar preferences in the social network and these relations are harmful for recommendation task. In other words, only those  social relations where social-pair users have high preference consistency can really facilitate the recommendation task. Therefore, we follow the principle of preference-guided social network denoising which requires evaluating the preference consistency of each social relation. Now we describe the process as follows.

Firstly, we employ LightGCN \cite{lightgcn} to perform multilayer convolutional operations for the learnable user embedding $\mathbf{e}_u$ and item embedding $\mathbf{e}_i$ on the interaction graph (all denoised interaction graph but the original interaction graph in the first training epoch) in order to learn the user preference embedding and item embedding:

\begin{equation}
\begin{aligned} 
& \mathbf{h}_u^{(l+1)}=\sum_{i \in \mathcal{N}_u} \frac{1}{\sqrt{\left|\mathcal{N}_u\right|} \sqrt{\left|\mathcal{N}_i\right|}} \mathbf{h}_i^{(l)}, \\
& \mathbf{h}_i^{(l+1)}=\sum_{u \in \mathcal{N}_i} \frac{1}{\sqrt{\left|\mathcal{N}_i\right|} \sqrt{\left|\mathcal{N}_u\right|}} \mathbf{h}_u^{(l)}.
\end{aligned}
\label{conv-lightgcn}
\vspace{-1em}
\end{equation}
where $\mathbf{h}_u^{(l)}$ and $\mathbf{h}_i^{(l)}$ represent the \textit{l}-th layer's embedding of user $\textit{u}$ and item $\textit{i}$, the initial embeddings $\mathbf{h}_u^{(0)}$ and $\mathbf{h}_i^{(0)}$ correspond to $\mathbf{e}_u$ and $\mathbf{e}_i$, respectively. Finally the user preference embedding $\mathbf{h}_u^{r}$ and item embedding $\mathbf{h}_i^{r}$ in interaction domain are obtained through average pooling operation on the embedding of all layers. Based on this, we compute the \textit{Preference Consistency} (\textit{PC}) between user $\textit{u}$ and user $\textit{v}$ of each social edge as follows: 
\begin{equation}
\begin{aligned}
    \textit{PC}\left(\textit{u}, \textit{v}\right)=\frac{1 + \cos\left(\mathbf{h}_{u}^r, \mathbf{h}_{v}^r\right)}{2}\cdot \exp\left(-\frac{||\mathbf{h}_{u}^r - \mathbf{h}_{v}^r||_2^2}{2\sigma^2}\right)
\end{aligned}
\label{pc_compute}
\end{equation}
where $\mathbf{h}_{u}^r$ and $\mathbf{h}_{v}^r$ represent the preference embeddings of user $\textit{u}$ and user $\textit{v}$, cos($\cdot$) represents consine similarity function. It is worth noting that we consider both direction consistency and magnitude consistency between preference embeddings to weigh the preference consistency more precisely, where $\sigma$ controls the weight of the Euclidean distance between $\mathbf{h}_{u}^r$ and $\mathbf{h}_{v}^r$ on consistency value, and normally we set it to 20. Eventually, the value of $\textit{PC}\left(\textit{u}, \textit{v}\right)$ is within [0,1]. A higher preference consistency means that the social edge is more trustworthy, and instead it is more likely a noisy edge. Therefore, in order to filter potential noisy edges, we set the threshold $\beta_s$ to remove the social edges with preference consistency scores less than the $\beta_s$ and keep the remaining ones, which makes the social network retain important social information while enabling the network structure more lightweight. Then the denoised social network is denoted by $\mathcal{G}_S'$. The denoising process is expressed as follows:
\begin{equation}
\begin{gathered}
\mathcal{G}_S' = \mathcal{G}_S \setminus \mathcal{N}_S, \\ 
\text{where} \ \mathcal{N}_S = \left\{(\textit{u}, \textit{v}) \in \mathcal{G}_S \ \big| \ \textit{PC}(\textit{u}, \textit{v}) < \beta_s \right\}
\end{gathered}
\label{denoise_s}
\end{equation}

\subsubsection{Social-enhanced interaction graph denoising}
Since we use the knowledge of interaction domain learned by noisy interaction graph to guide social network denoising, the denoised social network still contains noise. Therefore, we need to further denoise the interaction graph to better guide the next round of social network denoising. We believe that the noise in the interaction graph come from users' accident interactions or passive interactions due to social influence \cite{social_influence}. Therefore, we should consider whether the interacted items are truly compatible with the user. In fact, the formation of a reliable interaction graph follows the idea of user-based collaborative filtering \cite{collaborative_filtering}, i.e., user's preference on certain item can be inferenced by that of his/her similar users'. Hence, we draw support from the relatively cleaner social relations to derive user's true preference to maximally restore the true interaction graph. Specifically, We aggregate the preference embedding of user's first-order social neighbors, thereby obtaining social-enhanced user preference embedding $\mathbf{h}_u^{\text{enhanced}}$:
\begin{equation}
\begin{aligned}
\mathbf{h}_u^{\text{enhanced}} &= \text{AGG}_{\text{social}}\left(\left\{\mathbf{h}_v^r \ \big| \ \textit{v} \in \mathcal{S}_u \right\}\right)\\  
&= \sum_{v \in \mathcal{S}_u} \frac{1}{\sqrt{|\mathcal{S}_u| \cdot |\mathcal{S}_v|}} \mathbf{h}_v^r, \quad \forall \textit{u} \in \mathcal{U}
\end{aligned}
\label{agg}
\end{equation}
where $\mathcal{S}_u$ represents the set of user \textit{u}'s social neighbors. By doing this, the preferences within similar social groups are as consistent as possible, while the preferences between different social groups are as different as possible.
After obtaining the social-enhanced user preference embedding in interaction domain, we jointly use item embedding to compute the \textit{Interaction Compatibility (IC)} to determine whether the user with enhanced preference is compatible with their interacted items. Similarly, its calculation method is illustrated as:
\begin{equation}
\begin{aligned}
\textit{IC}\left(\textit{u}, \textit{i}\right)=\frac{1 + \cos\left(\mathbf{h}_u^{\text{enhanced}}, \mathbf{h}_i^r\right)}{2}\cdot \exp\left(-\frac{||\mathbf{h}_u^{\text{enhanced}} - \mathbf{h}_i^r||_2^2}{2\sigma^2}\right)
\end{aligned}
\label{ic_compute}
\end{equation}
Higher \textit{IC} value means that user \textit{u} is more compatible with item \textit{i}, and vice versa. Thus, we set a threshold $\beta_r$, for each edge in the interaction graph, if its interaction compatibility is lower than the $\beta_r$, it may be a potentially noisy edge thus should be deleted. The denoising process is expressed as follows:
\begin{equation}
\begin{gathered}
\mathcal{G}_R' = \mathcal{G}_R \setminus \mathcal{N}_R,\\
\text{where} \ \mathcal{N}_R = \left\{(\textit{u}, \textit{i}) \in \mathcal{G}_R \ \big| \ \textit{IC}(\textit{u}, \textit{i}) < \beta_r \right\}
\end{gathered}
\label{denoise_r}
\end{equation}
Through the above process, we reduce the noise in the original interaction graph, which is beneficial for the next round of social network denoising and provides cleaner interaction data for the recommendation task, thereby improving the recommendation performance of the model.

Through above two denoising segments, we achieve structure-level collaborative denoising which assure the denoising quality when there exists noise in both social network and interaction graph. At the same time, the social network truly plays a substantive role in providing supervisory signals for the interaction data. As the model iterates, $\mathcal{G}_R'$ and $\mathcal{G}_S'$ would get more and more trustworthy. 

\subsection{Embedding-space Collaborative Denoising}
\subsubsection{Noise cross-domain diffusion}
After obtaining the denoised social network $\mathcal{G}_S'$ and interaction graph $\mathcal{G}_R'$, we use parallel GNN encoders \cite{lightgcn} to learn the embeddings in interaction domain and social domain for the recommendation task:
\begin{equation}
\begin{aligned}
\mathbf{P}_{\mathcal{U}}^R, \mathbf{P}_{\mathcal{I}}^R = \text{GNN}^R(\mathbf{E}_{\mathcal{U}}, \mathbf{E}_{\mathcal{I}}, \mathcal{G}_R'),\\
\mathbf{P}^S_{\mathcal{U}} = \text{GNN}^S(\mathbf{E}_{\mathcal{U}}, \mathbf{E}_{\mathcal{I}}, \mathcal{G}'_S).
\end{aligned}
\label{ed_coder}
\end{equation}
where $\mathbf{P}_{\mathcal{U}}^R$ and $\mathbf{P}_{\mathcal{I}}^R$ are the user interaction embedding and item embedding matrixs, $\mathbf{P}_{\mathcal{U}}^S$ is the user social embedding matrix, and $\mathbf{E}_{\mathcal{U}}$ and $\mathbf{E}_{\mathcal{I}}$ are the matrix formulations of  $\mathbf{e}_u$ and $\mathbf{e}_i$.

However, $\mathbf{P}_{\mathcal{U}}^R$, $\mathbf{P}_{\mathcal{I}}^R$ and $\mathbf{P}_{\mathcal{U}}^S$ are containing noise. This problem arises from the following reason: The original interaction graph $\mathcal{G}_R$ and social network $\mathcal{G}_S$ contain noise while manual sturcture-level denoising process cannot completely eliminate the noise. This results in the phenomenon of noise cross-domain diffusion which would always accompany the structure denoising process. Specifically, the noise in interaction domain are diffused to social domain, and the noise in  social domain are diffused to  interaction domain. This leads to the embeddings learned by GNN based on the denoised interaction graph $\mathcal{G}_R'$ and social network $\mathcal{G}_S'$ still contain noise.

\begin{figure}[h]
    \centering
    \includegraphics[width=0.7\linewidth,height=5cm]{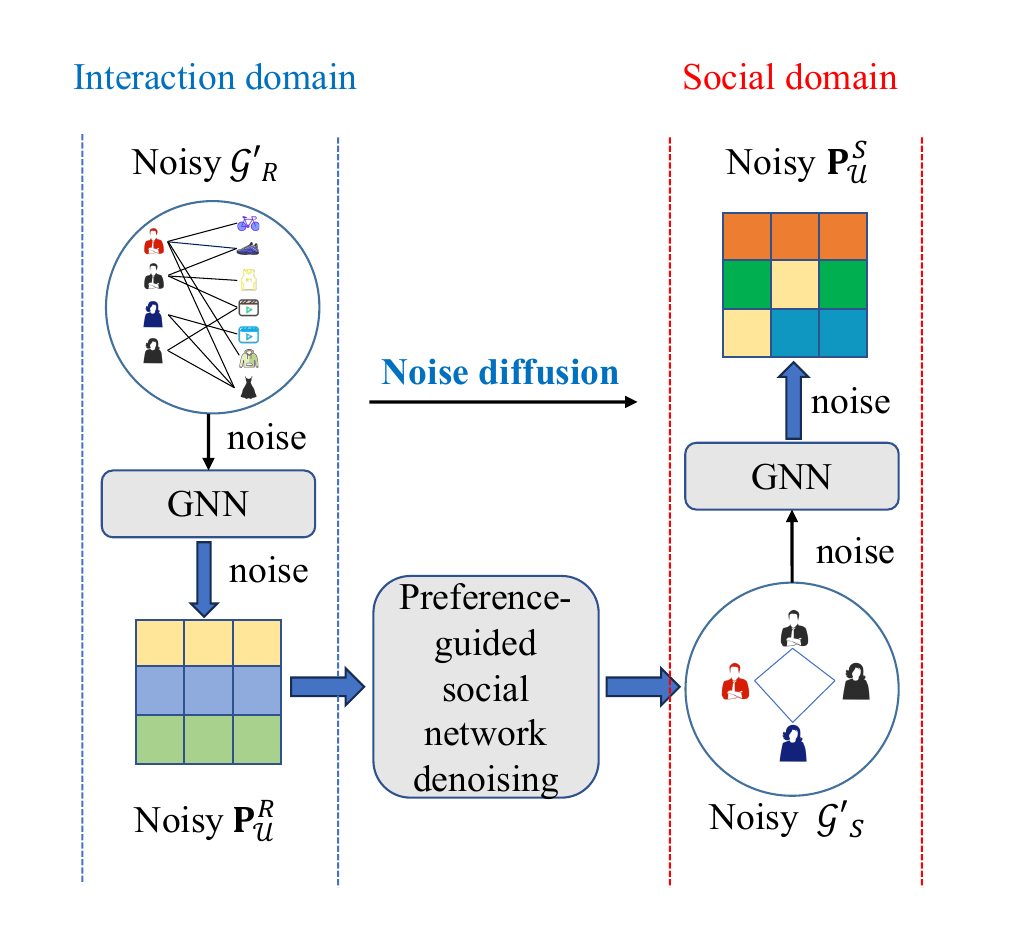}
    \caption{Noise diffusion from interaction domain to social domain}
    \label{fig:noise diffusion}
\end{figure}

Taking the noise diffusion from interaction domain to social domain as an example which is illustrated in Fig. \ref{fig:noise diffusion}, in the preference-guided social network denoising process, embeddings $\mathbf{h}_u^{r}$, $\mathbf{h}_i^{r}$ learned by GNN on the denoised but still noisy interaction graph $\mathcal{G}_R'$ naturally contain semantic noise. Subsequently, based on these embeddings, Eqs. (\ref{pc_compute})-(\ref{denoise_s}) are used for social network denoising, resulting in edges in $\mathcal{G}_S$ may be ignored for deletion or incorrectly deleted, thus the denoised social network $\mathcal{G}_S'$ still contains noise. Consequently, the $\mathbf{P}_{\mathcal{U}}^S$ learned based on $\mathcal{G}_S'$ contains noise. In summary, the noise in the interaction domain is diffused to the social domain. Similarly, social-enhanced interaction graph denoising process also diffuses the noise from the social domain to the interaction domain, causing $\mathbf{P}_{\mathcal{U}}^R$ and $\mathbf{P}_{\mathcal{I}}^R$ to contain noise. 

Therefore, we further perform denoising in the embedding space. We desire to restore the noise cross-domain diffusion process from the perspective of embedding space, then try to simulate the user and item embeddings uncontaminated by diffused noise in the form of dual-domain collaborative embedding perturbation, and finally perform embedding denoising with contrastive learning.

\subsubsection{Embedding Collaborative Perturbation}Specifically, above diffusion process we have illustrated can be simplified to that  $\mathbf{P}_{\mathcal{U}}^R$ symbolizing information of interaction domain and  $\mathbf{P}_{\mathcal{U}}^S$ symbolizing information of social domain  perturb each other. Therefore, for user social embeddings, we generate a kind of noise that satisfies the knowledge distribution of interaction domain:
                   
\begin{equation}
\Delta_S  = \text{sign}(\mathbf{P}_{\mathcal{U}}^S)\odot \frac{f_{shuffle}(\mathbf{P}_{\mathcal{U}}^R)}{||f_{shuffle}(\mathbf{P}_{\mathcal{U}}^R)||_2}
\label{noise_s}
\end{equation}
where $\text{sign}(\cdot)$ represents the symbol function which requires that $\mathbf{P}_{\mathcal{U}}^S$ and $\Delta_S$ are in the same hyperoctant, $\odot$ is the Hadama product, $f_{shuffle}$ \cite{RGD} means shuffling the rows of the $\mathbf{P}_{\mathcal{U}}^R$, aiming to simulate the uncertain false-positive connections among users generated in preference-guided social network denoising.

Subsequently, this noise is subtracted from the noisy user social embedding to simulate user social embedding uncontaminated by the noise from interaction domain  as follows:

\begin{equation}
\begin{aligned}
\mathbf{P}_{\mathcal{U}}^{S\prime}=\mathbf{P}_{\mathcal{U}}^S - \epsilon*\Delta_S
\end{aligned}    
\label{perturb_s}
\end{equation}
where the magnitude of final perturbation noise is numerically equivalent to $\epsilon$.


Similarly, for user interaction embeddings, we generate a kind of noise that satisfies the knowledge distribution of the social domain. Given that the social-enhanced interaction graph denoising process diffuses the noise from social domain to interaction domain, we subtract this noise from user interaction embedding to maximally simulate the raw user interaction embedding uncontaminated by the noise from social domain. It can be described as:

\begin{equation}
\begin{gathered}
\Delta_R = \text{sign}(\mathbf{P}_{\mathcal{U}}^R)\odot \frac{f_{shuffle}(\mathbf{P}_{\mathcal{U}}^S)}{||f_{shuffle}(\mathbf{P}_{\mathcal{U}}^S||_2},\\
\mathbf{P}_{\mathcal{U}}^{R\prime}=\mathbf{P}_{\mathcal{U}}^R - \epsilon*\Delta_R
\end{gathered}    
\label{perturb_r}
\end{equation}

Through above approach, we achieve dual-domain embedding collaborative perturbation, which explicitly defines the source and semantics of the perturbation noise. It can, to a certain extent, mitigate the noise cross-domain diffusion, thereby enabling the perturbed embeddings to maximally simulate the raw embeddings unaffected by noise diffusion.

In addition, for item embeddings, we use a similar method to generate noise that satisfy the item knowledge distribution, trying to mitigate the information propogation between items due to the noise in order to simulate the item embeddings unaffected by noise.

\begin{equation}
\begin{gathered}
\mathbf{P}_{\mathcal{I}}^{R\prime}=\mathbf{P}_{\mathcal{I}}^R - \epsilon*\mathbf{sign}(\mathbf{P}_{\mathcal{I}}^R)\odot \frac{f_{shuffle}(\mathbf{P}_{\mathcal{I}}^R)}{||f_{shuffle}(\mathbf{P}_{\mathcal{I}}^R)||_2}
\end{gathered}   
\label{perturb_i}
\end{equation}

Finally, above three embedding perturbation processes need to be repeated twice respectively to obtain positive samples $\mathbf{P}_{\mathcal{U}}^{S\prime}$, $\mathbf{P}_{\mathcal{U}}^{S\prime\prime}$ of user social embeddings, $\mathbf{P}_{\mathcal{U}}^{R\prime}$, $\mathbf{P}_{\mathcal{U}}^{R\prime\prime}$ of user interaction embeddings and $\mathbf{P}_{\mathcal{I}}^{R\prime}$, $\mathbf{P}_{\mathcal{I}}^{R\prime\prime}$ of item embeddings for contrastive learning.

\subsubsection{Contrastive Learning for embedding-space Denoising}After performing embedding perturbation and obtaining positive samples, we perform contrastive learning to denoise the original embeddings. Especially, we propose a novel contrastive learning loss function Anchor-InfoNCE (AC-InfoNCE) to improve InfoNCE \cite{infonce}, which set the original embedding as an anchor and simultaneously pull the anchor closer to its two positive samples together in the embedding space, while pushing apart embeddings from different users/items. InfoNCE ($\mathcal{L}_{\text{I}}$) and AC-InfoNCE ($\mathcal{L}_{\text{AC}}$) are illustrated as Eq. (\ref{ac-infonce_function}):
\begin{equation}
\begin{gathered}
\mathcal{L}_{\text{I}} = \sum_{i \in \mathcal{B}} -\log \left( \frac{\exp \left( s(\mathbf{h}_i^\prime, \mathbf{h}_i^{\prime\prime})/\tau \right)}{\sum_{j \in \mathcal{B}} \exp \left( {s(\mathbf{h}_i^\prime, \mathbf{h}_j^{\prime\prime})}/{\tau} \right)} \right)\\
\mathcal{L}_{\text{AC}} = \sum_{i\in\mathcal{B}} -\log \left( \frac{\exp \left( ({s(\mathbf{h}_i, \mathbf{h}_i^\prime) + s(\mathbf{h}_i, \mathbf{h}_i^{\prime\prime}))}/{2\tau} \right)}{\sum_{j\in\mathcal{B}} \exp \left( {s(\mathbf{h}_i^\prime, \mathbf{h}_j^{\prime\prime})}/{\tau} \right)} \right)\\
\label{ac-infonce_function}
\end{gathered}
\end{equation}
where $\mathbf{h}_i$ denotes the original embedding of user/item \textit{i}, $\mathbf{h}_i^\prime$, $\mathbf{h}_i^{\prime\prime}$ denotes two different augmented positive samples of $\mathbf{h}_i$, $\mathbf{h}_j^{\prime\prime}$ denotes the second positive sample of user/item \textit{j}, $s(\cdot)$ and $\tau$ represent similarity calculation operation which is cosine similarity here and temperature coefficient, respectively. 

\begin{figure}
    \centering
    \includegraphics[width=1\linewidth]{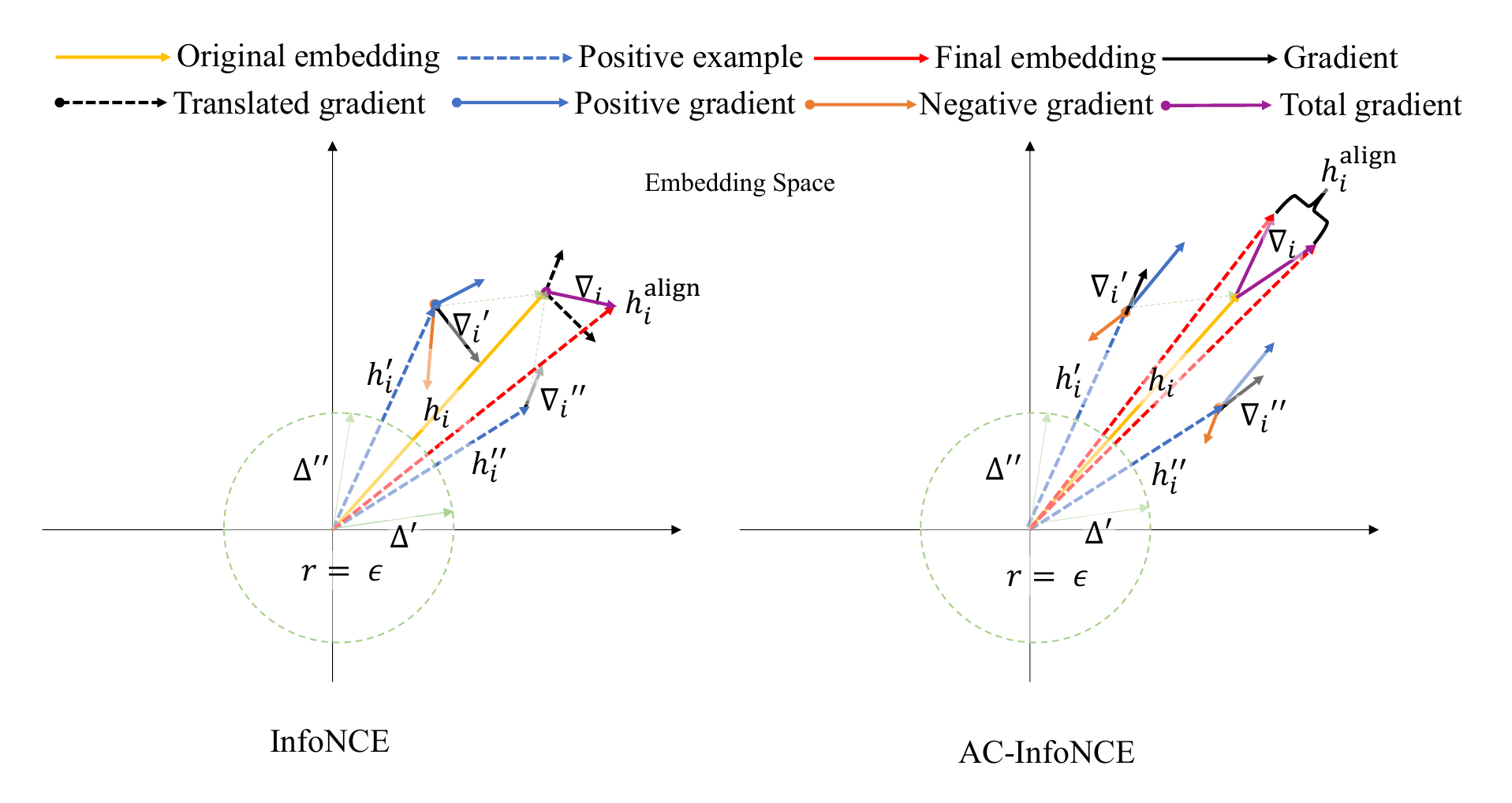}
    \caption{The process of embedding optimizing with InfoNCE and AC-InfoNCE.}
    \label{ac-infonce}
\end{figure}

AC-InfoNCE is proposed to allievate the gradient bias issue due to the uncertainty of embedding perturbation and the distribution of negative samples. Now we explain this issue and how AC-InfoNCE can effectively deal with it by combining numerical analysis (Eqs. (\ref{infonce_bias1})-(\ref{ac_bias2})) with geometrical representation (Fig. \ref{ac-infonce}). 

Eqs. (\ref{infonce_bias1})-(\ref{infonce_bias2}) respectively represent the gradient calculation process of InfoNCE loss with respect to $\mathbf{h}'_i$ and $\mathbf{h}''_i$ (i.e., $-\nabla'_i$ and $-\nabla''_i$), where $\tilde{\mathbf{h}_i^\prime}$ represent the $L_2$ norm of $\mathbf{h}_i^\prime$. We can see the fact that when $s(\mathbf{h}'_i, \mathbf{h}''_i)$ is very large while $s(\mathbf{h}'_i, \mathbf{h}''_j)$ and $s(\mathbf{h}''_i, \mathbf{h}''_j)$ are very small, $\nabla'_i$ and $\nabla''_i$ both approach 0, which is the driving force behind gradient update of InfoNCE. However, due to the uncertainty of data augmentation and the distribution of negative samples, there may be cases where $\mathbf{h}'_i$ has a high similarity with certain negative example $\mathbf{h}''_j$. The negative gradient may exacerbate the tendency for $\mathbf{h}'_i$ to approach $\mathbf{h}''_i$ to a certain extent, while $\mathbf{h}''_i$ still updates towards $\mathbf{h}'_i$ with a small scope, resulting in the final update direction of $\mathbf{h}_i$ ($\nabla_i$) being completely dominated by  $\nabla'_i$, in other words, it greatly biases $\mathbf{h}_i$ towards $\mathbf{h}''_i$. This is the so-called gradient bias issue, which would eventually lead to that the robustness of the finally optimized embedding $\mathbf{h}_i^{\text{align}}$ is vulnerable to embedding perturbation.

\begin{equation}
\begin{aligned}
   \frac{\partial \mathcal{L}_{\text{I}}}{\partial \mathbf{h}_{i}^{\prime}} &= \frac{1}{\tau}\left(\frac{\sum\limits_{j \in \mathcal{B}} \exp(s(\mathbf{h}_{i}^{\prime}, \mathbf{h}_{j}^{\prime\prime}) / \tau)) \cdot \frac{\partial s(\mathbf{h}_{i}^{\prime}, \mathbf{h}_{j}^{\prime\prime})}{\partial \mathbf{h}_{i}^{\prime}}}{\sum\limits_{j \in \mathcal{B}} \exp (s(\mathbf{h}_{i}^{\prime}, \mathbf{h}_{j}^{\prime\prime}) / \tau))}-\frac{\partial s(\mathbf{h}_{i}^{\prime}, \mathbf{h}_{i}^{\prime\prime})}{\partial \mathbf{h}_{i}^{\prime}}\right) \\
    &= \frac{1}{\tau} \left(\sum\limits_{j \in \mathcal{B}} \frac{\exp(s(\mathbf{h}_i^\prime, \mathbf{h}_j^{\prime\prime}) / \tau)}{\sum\limits_{k \in \mathcal{B}} \exp(s(\mathbf{h}_i^\prime, \mathbf{h}_k^{\prime\prime}) / \tau)} \frac{\mathbf{h}_j^{\prime\prime}}{\tilde{\mathbf{h}_i^\prime} \tilde{\mathbf{h}_j^{\prime\prime}}}- \frac{\mathbf{h}_i^{\prime\prime}}{\tilde{\mathbf{h}_i^\prime} \tilde{\mathbf{h}_i^{\prime\prime}}} \right)\\
    &= \frac{1}{\tau} \left(\mathbb{E}_{j \sim p(j|i)}\left[\frac{\mathbf{h}_j^{\prime\prime}}{\tilde{\mathbf{h}_i^\prime}\tilde{\mathbf{h}_j^{\prime\prime}}} \right] - \frac{\mathbf{h}_i^{\prime\prime}}{\tilde{\mathbf{h}_i^\prime}\tilde{\mathbf{h}_i^{\prime\prime}}} \right) \\
    & p(j|i) = \frac{\exp(s(\mathbf{h}_i^\prime, \mathbf{h}_j^{\prime\prime}) / \tau)}{\sum{k \in \mathcal{B}} \exp(s(\mathbf{h}_i^\prime, \mathbf{h}_k^{\prime\prime}) / \tau)}
    \end{aligned}
    \label{infonce_bias1}
\end{equation}


\begin{equation}
\begin{aligned}
\frac{\partial \mathcal{L}_{\text{I}}}{\partial \mathbf{h}_i^{\prime\prime}} &= \frac{1}{\tau} \left(p(i|i) \left[\frac{\mathbf{h}_i^{\prime}}{\tilde{\mathbf{h}_i^{\prime\prime}} \tilde{\mathbf{h}^{\prime}_i}}\right] - \frac{\mathbf{h}_i^{\prime}}{\tilde{\mathbf{h}_i^\prime} \tilde{\mathbf{h}_i^{\prime\prime}}}\right)·
\end{aligned}
\label{infonce_bias2}
\end{equation}

Based on this, we design the AC-InfoNCE, which sets $\mathbf{h}_i$ as an anchor point and prompts $\mathbf{h}_i^\prime$ and $\mathbf{h}_i^{\prime\prime}$ to converge to alignment with $\mathbf{h}_i$ at the same time.
According to Eq. (\ref{ac_bias1}), we can see that $\nabla_i$ essentially lies within the intermediate direction of $\mathbf{h}_i$ and $\mathbf{h}'_i$ and the intermediate direction of $\mathbf{h}_i$ and $\mathbf{h}''_i$. Even with the presence of negative sample $\mathbf{h}''_j$ which makes $s(\mathbf{h}'_i, \mathbf{h}''_j)$ high, it does not greatly affect the outcome. After embedding optimizing, $\mathbf{h}_i^{\text{align}}$ is obtained, which has only a minor deviation of $\mathbf{h}_i$ rather than significantly approaching $\mathbf{h}'_i$ or $\mathbf{h}''_i$, thereby ensuring that the final embeddings are robust. We can also draw this inclusion from Eq. (\ref{ac_bias2}) that both $\mathbf{h}'_i$ and $\mathbf{h}''_i$ are moving closer to $\mathbf{h}_i$, but ultimately $\nabla'_i$ and $\nabla''_i$ would not cross the direction of $\mathbf{h}_i$. This is the reason why $\mathbf{h}_i$ does not shift significantly in the end.
\begin{equation}
\begin{aligned}
\begin{split}
\frac{\partial \mathcal{L}_{\text{AC}}}{\partial \mathbf{h}_i} = &-\frac{1}{2\tau} \left(\frac{\mathbf{h}_i^\prime + \mathbf{h}_i}{\tilde{\mathbf{h}_i}\tilde{\mathbf{h}_i^\prime}} + \frac{\mathbf{h}_i^{\prime\prime} + \mathbf{h}_i}{\tilde{\mathbf{h}_i}\tilde{\mathbf{h}_i^{\prime\prime}}}\right)
\\
& +\frac{1}{2\tau}\left(\mathbb{E}_{j} \sim p(j|i) \left[\frac{2\mathbf{h}_j^{\prime\prime}}{\tilde{\mathbf{h}_i^\prime}\tilde{\mathbf{h}_j^{\prime\prime}}}\right]\right)
\end{split}
\end{aligned}
\label{ac_bias1}
\end{equation}

\begin{equation}
\begin{aligned}
\frac{\partial \mathcal{L}_{\text{AC}}}{\partial \mathbf{h}_i^\prime} &= -\frac{1}{2\tau}\left( \frac{\mathbf{h}_i}{\tilde{\mathbf{h}_i} \tilde{\mathbf{h}_i^\prime}} - \mathbb{E}_{j} \sim p(j|i)\left[\frac{2\mathbf{h}_j^{\prime\prime}}{\tilde{\mathbf{h}_i^\prime} \tilde{\mathbf{h}_j^{\prime\prime}}}\right]\right) \\
\frac{\partial \mathcal{L}_{\text{AC}}}{\partial \mathbf{h}_i^{\prime\prime}} &= -\frac{1}{2\tau}\left( \frac{\mathbf{h}_i}{\tilde{\mathbf{h}_i} \tilde{\mathbf{h}_i^{\prime\prime}}} -  p(i|i)\left[\frac{2\mathbf{h}_i^{\prime}}{\tilde{\mathbf{h}_i^\prime}\tilde{\mathbf{h}_i^{\prime\prime}}}\right]\right)
\end{aligned}    
\label{ac_bias2}
\end{equation}

Moreover, AC-InfoNCE incorporates the original embedding into contrastive learning, which is equivalent to treating it as a positive sample, expanding the number of positive samples for contrastive learning, thus alleviating the data sparsity problem and further realizing the denoising effect.

\subsection{Joint Training}DCDSR utilizes multi-task joint training to tune the model. The entire model incorporates the losses of recommendation task and contrastive learning task to guide the model in learning the user and item embeddings. Given the user interaction embedding and item embedding, we predict the likelihood of user \textit{u} interacting with item \textit{i}, formulated as:
\begin{equation}
\hat{y}_{ui}=\mathbf{p}_u^{r\text{T}}\mathbf{p}_i
\label{inner_product}
\end{equation}

Based on this, we minimize the following BPR \cite{2012bpr} loss to optimize the main recommendation task:
\begin{equation}
\mathcal{L}_{\text {BPR }}=\sum_{(u, i, j) \in O}-\log \sigma\left(\hat{y}_{u, i}-\hat{y}_{u, j}\right)
\label{bpr_compute}
\end{equation}
where $\sigma(\cdot)$ is the sigmoid function, $\mathcal{O}=\{(u, i, j) \mid(u, i) \in \mathcal{O}^{+},(u, j) \in \mathcal{O}^{-}\}.$ Here, $\mathcal{O}^{+}$ is the observed pairwise training data, and $\mathcal{O}^{-}$ denotes the unobserved interactions. The contrastive learning task consists of CL losses $\mathcal{L}_{\text{CL}}^r$ for user interaction embeddings , $\mathcal{L}_{\text{CL}}^s$ for user social embeddings, and $\mathcal{L}_{\text{CL}}^i$ for item embeddings, as follows:

\begin{equation}
\begin{aligned}
\mathcal{L}_{\text{CL}}^r = \sum_{u \in \mathcal{B}} -\log \left( \frac{\exp \left( ({s(\mathbf{p}_u^r, \mathbf{p}_u^{r\prime}) + s(\mathbf{p}_u^r, \mathbf{p}_u^{r\prime\prime}))}/{2\tau} \right)}{\sum_{v \in \mathcal{B}} \exp \left( {s(\mathbf{p}_u^{r\prime}, \mathbf{p}_v^{r\prime\prime})}/{\tau} \right)} \right)\\
\mathcal{L}_{\text{CL}}^s = \sum_{u \in \mathcal{B}} -\log \left( \frac{\exp \left( ({s(\mathbf{p}_u^s, \mathbf{p}_u^{s\prime}) + s(\mathbf{p}_u^s, \mathbf{p}_u^{s\prime\prime}))}/{2\tau} \right)}{\sum_{v \in \mathcal{B}} \exp \left( {s(\mathbf{p}_u^{s\prime}, \mathbf{p}_v^{s\prime\prime})}/{\tau} \right)} \right)\\
\mathcal{L}_{\text{CL}}^i = \sum_{i \in \mathcal{B}} -\log \left( \frac{\exp \left( ({s(\mathbf{p}_i^r, \mathbf{p}_i^{r\prime}) + s(\mathbf{p}_i^r, \mathbf{p}_i^{r\prime\prime}))}/{2\tau} \right)}{\sum_{j \in \mathcal{B}} \exp \left( {s(\mathbf{p}_i^{r\prime}, \mathbf{p}_j^{r\prime\prime})}/{\tau} \right)} \right)
\end{aligned}
\label{cl_loss_compute}
\end{equation}

Finally, the loss of the recommendation task and the losses of the contrastive learning task are combined to obtain the total loss of DCDSR as follows:

\begin{equation}
\begin{split}
\mathcal{L}_{\text{DCDSR}}=\mathcal{L}_{\text{BPR}}&+\lambda_1\mathcal{L}_{\text{CL}}^r+\lambda_{2\ }\mathcal{L}_{\text{CL}}^s+\lambda_{3\ }\mathcal{L}_{\text{CL}}^i
\\
&+\lambda_{\text{reg}}\left(||\mathbf{E}_{\mathcal{U}}||_F^2+||\mathbf{E}_{\mathcal{I}}||_F^2\right)
\end{split}
\label{total_loss}
\end{equation}
where $\lambda_1$, $\lambda_2$ and $\lambda_3$ are the coefficients controlling the effects of three CL loss respectively, and $\lambda_{\text{reg}}$ represents the regularization weight for the user/item embedding.
The whole training process of DCDSR is shown in Algorithm \ref{Algorithm 1}.
\begin{algorithm}[t]
    \caption{The overall training process of DCDSR}
    \label{Algorithm 1}
    
    \KwIn{The original interaction graph $\mathcal{G}_R$  and social network $\mathcal{G}_S$, hyperparameters $\beta_r$, $\beta_s$, $\lambda_1$, $\lambda_2$, $\lambda_3$;}
    \KwOut{ The user embeddings $\mathbf{E}_{\mathcal{U}}$ and item embeddings $\mathbf{E}_{\mathcal{I}}$;}
            \begin{algorithmic}[1]
            \STATE Initialize parameters $\mathbf{E}_{\mathcal{U}}$ and $\mathbf{E}_{\mathcal{I}}$ with the default Xavier distribution;\\
            \STATE \textbf{while} DCDSR not converge \textbf{do}\\
            
            \STATE \quad Reconstruct a denoised social network $\mathcal{G}_\mathcal{S}'$ based on original social network $\mathcal{G}_{S} $ with Eqs. (\ref{conv-lightgcn})-(\ref{denoise_s});\\
            \STATE \quad Reconstruct a denoised interaction graph $\mathcal{G}_{R}'$ based on denoised social network $\mathcal{G}_{S}'$ with Eqs. (\ref{agg})-(\ref{denoise_r});\\
            \STATE \quad Sample a mini-batch of user $\mathcal{B}\in\mathcal{U}$;\\
            \STATE \quad \textbf{for} \textit{u} $\in$ $\mathcal{B}$ \textbf{do}\\
            \STATE \quad\quad Obtain user social embeddings $\mathbf{P}_{\mathcal{U}}^{R}$, item embeddings $\mathbf{P}_{\mathcal{I}}^{R}$ and user social embeddings $\mathbf{P}_{\mathcal{U}}^{S}$ for each user \textit{u} and item \textit{i} based on $\mathcal{G}_{R}'$ and $\mathcal{G}_{S}'$ with Eq. (\ref{ed_coder});\\
            \STATE \quad\quad Calculate BPR loss $\mathcal{L}_\text{BPR}$ with Eqs. (\ref{inner_product})-(\ref{bpr_compute});\\
            \STATE \quad\quad Obtain perturbed embeddings $\mathbf{P}_{\mathcal{U}}^{R\prime}$, $\mathbf{P}_{\mathcal{U}}^{R\prime\prime}$, $\mathbf{P}_{\mathcal{U}}^{S\prime}$, $\mathbf{P}_{\mathcal{U}}^{S\prime\prime}$, $\mathbf{P}_{\mathcal{I}}^{R\prime}$, and $\mathbf{P}_{\mathcal{I}}^{R\prime\prime}$ based on $\mathbf{P}_{\mathcal{U}}^{R}$, $\mathbf{P}_{\mathcal{U}}^{S}$, and $\mathbf{P}_{\mathcal{I}}^{R}$ with Eqs. (\ref{noise_s})-(\ref{perturb_i});\\
            \STATE \quad\quad Calculate AC-InfoNCE losses $\mathcal{L}_{\text{CL}}^r$,$\mathcal{L}_{\text{CL}}^s$, and $\mathcal{L}_{\text{CL}}^i$ for contrastive learning, according to Eq. (\ref{cl_loss_compute});\\
            \STATE \quad\quad Calculate joint learning loss $\mathcal{L}_{\text{DCDSR}}$ with Eq. (\ref{total_loss}); \\
            \STATE \quad\quad Backpropagation and update parameters $\mathbf{E}_{\mathcal{U}}$ and $\mathbf{E}_{\mathcal{I}}$ with $\mathcal{L}_{\text{DCDSR}}$ ; \\
            \STATE \quad \textbf{end for};\\
            \STATE \textbf{end while};\\
            \STATE \textbf{return} parameters $\mathbf{E}_{\mathcal{U}}$ and $\mathbf{E}_{\mathcal{I}}$;\\
            \end{algorithmic}
\end{algorithm}

\subsection{Discussion}
\subsubsection{Novelty and Difference}
In this section, we compare DCDSR with various denoising GSR and GCL-based GSR models to highlight the novelty of DCDSR.

Firstly, DCDSR is a pioneering denoising GSR model that collaboratively performs denoising on social domain and interaction domain, moreover, it excutes denoising at both the structure level and the embedding space. In contrast, DSL \cite{dsl} only performs social domain denoising through adaptive alignment at the embedding space. GDMSR \cite{gdmsr} jointly denoises the social domain at the structure level and embedding space. It can be seen that these models do not consider denoising the interaction domain noise, and some of them only perform denoising at a single perspective.

In addition, the embedding-space collaborative denoising module of DCDSR incorporates graph contrastive learning. In terms of data augmentation method, unlike the GA used in SEPT \cite{sept} and DcRec \cite{dcrec}, DCDSR adopts FA, which is more efficient and avoids the loss of important graph structure information during the GA process. At the same time, compared to general FA method \cite{simgcl} which performs embedding perturbation at each layer of GNN, DCDSR only performs embedding perturbation at the last layer of GNN. Furthermore, the noise used for embedding perturbation in DCDSR is sampled from the interaction/social domain knowledge distribution, rather than the random Gaussian noise \cite{simgcl}. This approach restores the process of noise cross-domain diffusion in social recommendation and achieves embedding-space denoising with better explainability.

\subsubsection{Complexity Analysis}
In this section, we analyze the theoretical temporal
complexity of DCDSR. Table \ref{complexity} provides a
comparison between DCDSR and some state-of-the-art graph-based recommendation models in terms of time complexity. Some models introduce GCL while others are designed for denoising task. Therefore, temporal complexity analysis of GNN operation and structure-level denoising (SD) are given and whether they performs emebdding-space denoising (ED) is provided. The numbers of edges and vertices of the interaction graph
are $|E|$ and $|V|$, while the number of edges in the social
network is represented by $|S|$. In the encoder, $L$ represents the number of GNN layers, $d$ denotes the embedding size, $s$ is the number of epochs required for training, $B$ indicates the batch size, and $\rho$ corresponds to the
edge retention rate in GA. Furthermore,
$A_f$, As represent the friend and share views engendered
by the SEPT method, respectively, while $H_s$, $H_j$, $H_p$ are
the triad of convolutional channels for social, joint, and
purchase within MHCN. \textit{D} denotes to a social network reconstruction every \textit{D} epochs in GDMSR. For DCDSR, the time complexity for one iteration is consist of GNN encoding which consumes
$O((|E| + |S|)Ld * |E|/B)$ and SD for social network and interaction graph and  which totally consume $O(|E| + |S|)$. Therefore, after \textit{s} iterations, the total
time complexity becomes $O((|E| + |S|) Lds|E|/B + (|E| + |S|)s)$.

\begin{table}[t]

\caption{Temporal Complexity Analysis.}
\centering
\scalebox{0.7}{
\begin{tabular}{l|c|c|c}
\hline & GNN & GD & ED  \\
\hline \hline NGCF & $O((|E|+|V|) L d s|E| / B)$ & $\times$ & $\times$  \\
LightGCN & $O(|E| L d s|E| / B)$ & $\times$ & $\times$   \\
DiffNet & $O((|S| L+|E|) d s|E| / B)$ & $\times$ & $\times$ \\
$S^2$-MHCN & $O\left(\left(\left|H_s\right|+\left|H_j\right|+\left|H_p\right|+|E|\right) L d s|E| / B\right)$ & $\times$ & $\times$ \\
SEPT & $O\left(\left(\left|A_f\right|+\left|A_s\right|+|E|+\rho_1(|E \cup S|)\right) L d s|E| / B\right)$ & $\times$ & $\times$ \\
DESIGN & $O(2(|E|+|S|) L d s|E| / B)$ & $\times$ & $\times$ \\
DcRec & $O\left(\left(\left(1+2 \rho_1\right)|E|+2 \rho_2|S|\right) L d s|E| / B\right)$ & $\times$ & $\times$ \\
DSL & $O((|E|+|S|) L d s|E| / B)$ & $\times$ & $\checkmark$ \\
GDMSR & $O((|E|+|S|) L d s|E| / B)$ & $O(|S|s/D)$ & $\times$ \\
DCDSR & $O((|E|+|S|) L d s|E| / B)$ & $O((|E| + |S|)s)$ & $\checkmark$ \\
\hline
\end{tabular}}
\label{complexity}
\end{table}

\section{Experiments}
We conduct experiments on three real-world datasets to validate the superiority of DCDSR compared to some state-of-the-art models and answer the following questions:\\
$\bullet$ RQ1: How does DCDSR perform comparing to these GSR and denoising social recommendation models?\\
$\bullet$ RQ2: Do each key component and method of DCDSR contribute to the model?\\
$\bullet$ RQ3: How about the noise-resistance of DCDSR compare to traditional GSR and denoising social recommendation models in the face of different proportions of noise attack?\\
$\bullet$ RQ4: How does the performance of DCDSR change as the values of the model hyperparameters change?

\subsection{Datesets}
In this paper, we adopt six benchmark datasets Douban-book \cite{sept}, Yelp\cite{yin2019social}, Douban-movie \cite{mhcn}, Ciao \cite{gdmsr}, Yelp2 \cite{gdmsr} and Douban \cite{gdmsr} to evaluate DCDSR, all of which are publicly available and cover different domains with various data sizes and densities. Table \ref{datasets} lists the details of these datasets. In order to be consistent with previous work, the interaction data from all the datasets are transformed into implicit feedback \cite{NGCF,lightgcn,sept} and divided into the training set and the test set in a ratio of 8:2.

\begin{table}
\centering
\caption{Statistics of datasets.}
\begin{tabular}{c|cccc}
\hline Dataset & \#User & \#Item & \#Interaction & \#Relation\\
\hline \hline Douban-Book & 13,024 & 22,347 & 792,062 & 169,150\\
Yelp & 19,539 & 21,266 & 450,884 & 363,672 \\
Douban-movie & 2,848 & 39,586 & 894,887 & 35,770\\
Ciao & 7,335 & 17,867 & 140,628 & 111,679  \\
Yelp2 & 32,827 & 59,972 & 598,121 & 964,510  \\
Douban & 2,669 & 15,940 & 535,210 & 32,705  \\
\hline
\end{tabular}
\label{datasets}
\end{table}

\begin{table*}[h]
\caption{Overall performance comparison. The model that performs best on each dataset and metric is bolded, whereas the best beseline is underlined. ‘Improve. \%’ indicates the relative improvement of DCDSR over the best baseline. R@10, N@10, R@20 and N@20 represent Recall@10, NDCG@10, Recall@20 and NDCG@20, respectively. Symbol * denotes that the improvement is significant with a p-value $\textless$ 0.05 based on a two-tailed paired t-test.}
\renewcommand\arraystretch{1.5}
\resizebox{\linewidth}{!}{
\begin{tabular}{cc|ccccc|ccc|cc|c}
\hline Datasets & Metrics & NGCF & LightGCN & DiffNet & MHCN & DESIGN & $S^2$-MHCN  & SEPT & DcRec & DSL & \textbf{DCDSR} & Improve. \% \\
\hline\hline \multirow{4}{*}{\begin{tabular}{c} 
Douban- \\
Book
\end{tabular}} &  R@10 & 0.0813 & 0.0900 & 0.0845 & 0.1011 & 0.0945 & 0.1063
 & 0.1058 & $\underline{0.1085}$ & 0.1005 & $\mathbf{0.1252}^*$ & $+15.39\%$ \\
 & N@10 & 0.1015 & 0.1238 & 0.1068 & 0.1154 & 0.1255
 & 0.1385 & 0.1373
 & \underline{0.1404}
 & 0.1317 & $\mathbf{0.1689}^*$ & $+16.64 \%$ \\
 & R@20 & 0.1261 & 0.1376 & 0.1272 & 0.1504 & 0.1416
 & 0.1559
 & 0.1545 & $\underline{0.1597}$ & 0.1503 & $\mathbf{0.1794}^*$ & $+12.34 \%$ \\
 & $\mathrm{N} @ 20$ & 0.1112 & 0.1302 & 0.1144 & 0.1408 & 0.1300
 & 0.1461
 & 0.1432 & \underline{0.1466}
 & 0.1376 & $\mathbf{0.1726}^*$ & $+17.18 \%$ \\
\hline \multirow{4}{*}{ Yelp } & R@10 & 0.0554 & 0.0643 & 0.0535 & 0.0687 & 0.0703
 & $\underline{0.0789}$
 & 0.0746  & 0.0761 & 0.0727 & $\mathbf{0.0871}^*$ & $+10.39 \%$ \\
 & $\mathrm{N} @ 10$ & 0.0419 & 0.0512 & 0.0406 & 0.0531 & 0.0535
 & $\underline{0.0606}$
 & 0.0578 & 0.0585 & 0.0551 & $\mathbf{0.0670}^*$ & $+10.56 \%$ \\
 & R@20 & 0.0903 & 0.1048 & 0.0873 & 0.1087 & 0.1113
 & $\underline{0.1207}$
 & 0.1126 & 0.1205 & 0.1164 & $\mathbf{0.1346}*$ & $+11.52 \%$ \\
 & $\mathrm{N@20}$ & 0.0533 & 0.0641 & 0.0516 & 0.0657 & 0.0662
 & $\underline{0.0740}$
 & 0.0676 & 0.0723 & 0.0687 & $\mathbf{0.0820}^*$ & $+10.81 \%$ \\
\hline \multirow{4}{*}{\begin{tabular}{c} 
Douban- \\
Movie
\end{tabular}} & R@10 & 0.0507 & 0.0533 & 0.0473 & 0.0541 & 0.05513
 & 0.0554
 & $\underline{0.0574}$ & 0.0573 & 0.0567 & $\mathbf{0.0661}^*$ & $+15.16 \%$ \\
 & N@10 & 0.2315 & 0.2491 & 0.2049 & 0.2409 & 0.2429
 & 0.2572
 & $\underline{0.2672}$ & 0.2508 & 0.2470 & $\mathbf{0.2862}^*$ & $+7.11 \%$ \\
 & R@20 & 0.0830 & 0.0883 & 0.0816 & 0.0909 & 0.0934 & 0.0926
 & 0.0932 & $\underline{0.0947}$ & 0.0923 & $\mathbf{0.1116}^*$ & $+17.85 \%$ \\
 & $\mathrm{N} @ 20$ & 0.2161 & 0.2331 & 0.1944 & 0.2251 & 0.2324 & 0.2375
 & $\underline{0.2466}$ & 0.2333 & 0.2287 & $\mathbf{0.2663}^*$ & $+7.99 \%$ \\
\hline
\end{tabular}}
    \label{overall comparsion}
\end{table*}

\begin{table*}[t]
\caption{Performance comparison between GDMSR and DCDSR. GDMSR$_D$ and GDMSR$_M$ reprenset employing GDMSR based on DiffNet++ \cite{diffnet++} and MHCN, respectively.}
\renewcommand\arraystretch{1.5}
\resizebox{\linewidth}{!}{
\begin{tabular}{lcccccc}
\hline & \multicolumn{2}{c}{ Ciao } & \multicolumn{2}{c}{ Yelp2 } & \multicolumn{2}{c}{ Douban } \\
\cline { 2 - 7 } &  Recall@3 & NDCG@3 &  Recall@3 & NDCG@3 &  Recall@3 & NDCG@3  \\
\hline GDMSR$_D$ [41]  & \underline{0.1244} & \underline{0.2821}   & 0.3291 & 0.6102  & 0.2540 & 0.5297\\
GDMSR$_M$ [3]  & 0.1138 & 0.2632  &  \underline{0.3405} & \underline{0.6434}  & \underline{0.3496} & \underline{0.6137}\\
\hline DCDSR [31]  & $\mathbf{0.1426^*}(\mathbf{+14.63 \%})$ & $\mathbf{0.3084^*}(\mathbf{+9.32 \%)}$ &  $\mathbf{0.4517^*}(\mathbf{+32.66 \%})$ & $\mathbf{0.6471^*}(\mathbf{+0.57 \%})$ & $\mathbf{0.4561^*}(\mathbf{+30.46 \%})$ & $\mathbf{0.7003^*}(\mathbf{+14.11 \%})$\\
\hline
\end{tabular}}
\label{gmdsrvsdcdsr}
\end{table*}

\subsection{Baselines}To validate the effectiveness of the proposed model and answer the questions raised above, we compare DCDSR with the following baselines:\\
$\bullet$ \textbf{NGCF} \cite{NGCF}: This model utilizes standard GCN to explore complex relationships between users and items, boosting neighbor propagation through element-wise multiplication.\\
$\bullet$ \textbf{LightGCN} \cite{lightgcn}: This model streamlines GCN for recommendation by eliminating nonlinear activations, feature transformations, and self-connections.\\
$\bullet$ \textbf{DiffNet} \cite{diffnet}: This model portrays social information diffusion process in social networks for GSR.\\
$\bullet$ \textbf{MHCN} \cite{mhcn}: This model uses multi-channel hypergraph convolutional networks to model complex user relationships for GSR.\\
$\bullet$ \textbf{DESIGN} \cite{design}: This GSR method constructs models of SocialGCN, RatingGCN, and TwinGCN to respectively model users and items, and employs knowledge distillation to collaboratively boost recommendation performance.\\
$\bullet$ $\boldsymbol{S^2}$-$\textbf{MHCN}$ \cite{mhcn}: This model introduces GCL into \textbf{MHCN}.\\
$\bullet$ \textbf{SEPT} \cite{sept}: This leading GSR model constructs various social network perspectives for GCL to encode multiple user embeddings , providing supervisory signals to the main task.\\
$\bullet$ \textbf{DcRec} \cite{dcrec}: This GSR model constructs augmented graphs to learn different perspectives of node embeddings from both interaction and social domains to provide supervision signals.\\
$\bullet$ \textbf{DSL} \cite{dsl}: This method is a denoising GSR model that enables denoised cross-view alignment for similarities between different users .\\
$\bullet$ \textbf{GDMSR} \cite{gdmsr}: This method is a denoising framework for GSR which is preference-guided to model social relation confidence and benefits user preference learning in return by providing a denoised but more informative social graph for recommendation models.

\subsection{Parameter Settings} We implement our DCDSR in PyTorch \footnote{https://pytorch.org/}. All experiments are conducted on a single Linux server with Xeon(R) Gold 6133 CPU, 20G RAM, and 1 NVIDIA GeForce RTX 3090 GPU. All models are optimized using Adam \cite{adam} with parameters initialized by the Xavier method \cite{xavier}. For fair comparisons, the embedding size is set to 50 for all the baselines except GMDSR, where we set it to 8 according to the original paper. The mini-batch size and number of GCN layers is fixed to 2048 and 2 for all models. Following the original papers' suggested settings, we perform a grid search to select the optimal hyper-parameters for all models. For DCDSR, we tune the structure-level collaborative denoising threshold $\beta_s$ and $\beta_r$ in the range of [0.5, 0.6, 0.7, 0.8, 0.9] and [0.3, 0.35, 0.4, 0.45, 0.5], respectively. The proportional coefficients of contrastive losses in three different views are all tuned in the range of [0.1, 0.2, 0.3, 0.4]. The temperature parameter for AC-InfoNCE and the magnitude of perturbation noise $\epsilon$ are tuned within [0.1, 0.2, 0.5, 1.0] and [0, 0.1,..., 1.0], respectively. 

\subsection{Evaluation Protocols}To assess the performance of Top-N recommendation, in regular baselines, we employ two commonly used evaluation metrics Recall and NDCG which are computed by the all-ranking protocol \cite{lightgcn, sgl}. And in GDMSR, real-plus-N \cite{gdmsr} is adopted to compute these metrics which randomly selects 100 uninteracted items for each user and ranks them together with the positive test samples.

\subsection{Performance Comparison (RQ1)}

As Table \ref{overall comparsion} and Table \ref{gmdsrvsdcdsr} show, Overall, on each datasets, DCDSR achieves the best performance on every metric, achieving at average of more than 10\% improvement over all baseline models. Specifically, on Douban-Book, Yelp, and Douban-Movie, DCDSR achieves 12.27\%, 15.95\%, and 10.69\% improvement on Recall@20, and 17.74\%, 10.63\%, and 10.54\% improvement on NDCG@20. On Ciao, Yelp2 and Douban, DCDSR achieves 14.63\%, 32.66\%, and 30.46\% improvement on Recall@3, and 9.32\%, 0.57\%, and 14.11\% improvement on NDCG@3 compared to GMDSR. This is a strong proof of the significant effectiveness of DCDSR and its great universality to different data environments. We attribute these improvemens to the following reasons:

(1) Adhering to the dual-domain collaborative denoising paradigm ensures that the social network denoising task is less vulnerable to the noise in original interaction data. This approach not only enhances the quality of social network but also optimizes interaction data, thereby directly facilitating recommendation task. Building on preference-guided social network denoising, the denoised social network further serves to purify the interaction graph through knowledge enhancement, thus better exploiting the supervisory role of social network in recommendation task.

(2) The embedding-space collaborative denoising module to some extent restores the noise cross-domain diffusion process. It simulates the raw embeddings unaffected by noise cross-domain diffusion through dual-domain embedding collaborative perturbations. Finally, these perturbed embeddings, along with the original embeddings, are used for contrastive learning, which denoises the embeddings and successfully mitigates the noise cross-domain diffusion issue. In addition, user social embeddings and user interaction embeddings are united by the contrastive learning task thus providing additional supervised signals for the recommendation task, bridging the information barriers between social and interaction domains.




Combining all these baselines, GCL significantly improves the performance of GSR models. Taking DcRec as an example, it improves the performance of GSR by decoupling the user behavior into the representation of collaborative and social domains, and adopting the decoupled contrastive learning mechanism to transfer knowledge between the two domains. But it has the following drawbacks compared to DCDSR: (1) It has not considered the noise issue in social recommendation. (2) It generates different views for contrastive learning by random graph augmentation, which may lose important edges in the original graph/network structure, thus leading to degradation of recommendation performance.

Further, we compare DSL and GDMSR with DCDSR, respectively. As it can be seen, DCDSR significantly outperforms DSL on all  datasets due to the fact that DSL implicitly reduces noise by aligning the similarity of user-user pairs in the interaction domain with that in the social domain, and computes the BPR loss for social domain embeddings, which allow similar users have similar preferences. However, it has several drawbacks: (1) It overlooks the noise in interaction domain, thus failing to ensure the authenticity of the learned user preferences. (2) BPR loss for social network link prediction based on noisy social network actually defeats the purpose of preference-guided denoising for social domain due to the information propogation between noisy social relations and interferes with the process of adaptive alignment of user representations in interaction and social domains. 

Similarly, GDMSR computes the confidence of social relations and performs link prediction task for self-correction by utilizing the item embeddings in the interaction domain to represent user social embeddings. Additionally, it adaptively denoises the structure of the social network at regular intervals during the training process, achieving a certain level of noise reduction. However, it still overlooks the noise within the interaction domain, which compromises the overall denoising effect and the performance of the recommendation task. 

In contrast, DCDSR not only utilizes the knowledge from interaction domain to guide the social network denoising, but also utilizes the social network to enhance the knowledge of interaction domain  for denoising in turn and repeat above process for adaptive optimizing, which realizes the bidirectional supervision of interaction and social domains. Meanwhile, the GNN-encoded user social embeddings are only served for the contrastive learning task and will not interfere with the recommendation task.


\subsection{Ablation Study (RQ2)}In this section, we evaluate the effectiveness of the major components of DCDSR  and further analyze the dual-domain embedding collaborative perturbation and AC-InfoNCE.
\subsubsection{\textbf{Effectiveness of Major Components}}

\begin{figure}
    \centering
    \includegraphics[width=\linewidth]{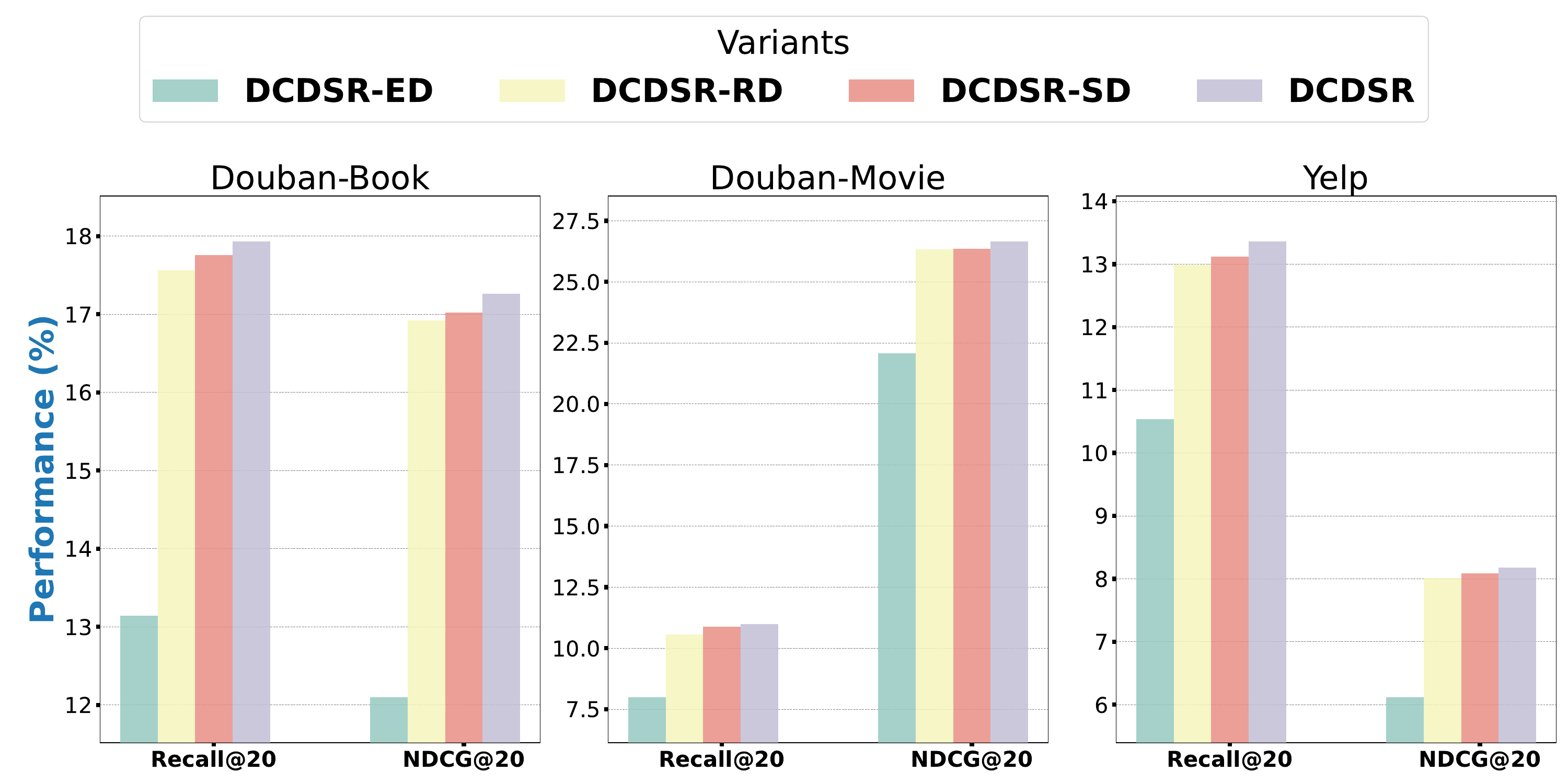}
    \caption{Contributions of each major component in DCDSR.}
    \label{ablation_graph}
\end{figure}

In order to test the effectiveness of each important component of DCDSR, we performed ablation experiments on them. Specifically, we compare the performance between DCDSR and the following ablation variants of DCDSR: (1) "DCDSR-RD": removing denoising for interaction domain at both structure level and embedding space while only adopt preference-guided social domain denoising. (2) "DCDSR-SD": removing the structure-level collaborative denoising module used to produce cleaner social network and interaction graph. (3) "DCDSR-ED": removing the embedding-space collaborative denoising module which restores the noise cross-domain diffusion process to learn denoised user and item embeddings. The results are shown in Fig. \ref{ablation_graph}.   \\

By comparison, we find that DCDSR performs better than other all ablation variants. Moreover, by comparing each variant we can draw the following conclusions: 
(1) The performance of DCDSR-RD shows a significant decline compared to DCDSR, which demonstrates that denoising for social domain is severely affected when there exists noise in interaction domain. This proves the correctness of our motivation and necessity of our work.
(2) DCDSR-SD exhibits a notable performance decrease compared to DCDSR, indicating that ignoring structure-level collaborative denoising and employing noisy social network and interaction graph for GNN encoding allows the recommendation task to be greatly disturbed by structural noise.
(3) DCDSR-ED experiences a substantial performance drop compared to DCDSR, suggesting that the embedding-space collaborative denoising module plays a crucial role in the model. It mitigates the adverse effects of noise cross-domain diffusion. In other words, It can  discriminate and cut down the noise that escape during structure-level collaborative denoising.
\subsubsection{\textbf{Embedding Collaborative Perturbation vs Embedding Random Perturbation}}
Embedding \textbf{R}andom \textbf{P}erturbation (\textbf{RP}) is a common method of data augmentation for GCL, where perturbed embeddings are obtained by adding noise sample from a certain distribution such as Gaussian distribution \cite{simgcl} to the original embeddings. However, we argue that employing the random noise used for embedding perturbation does not have a specific semantic meaning and ignores the source of the noise, making the data augmentation imprecise. In contrast, the dual-domain embedding \textbf{C}ollaborative  \textbf{P}erturbation (\textbf{CP}) proposed by DCDSR specifies the source and semantics of the noise and imaginatively restores the noise cross-domain diffusion  generated in structure-level collaborative denoising module, thus improving the explainability of data augmentation, which is conducive to learning more accurate and robust embeddings. As shown in Table \ref{CP VS RP}, $\text{DCDSR}_{CP}$ performs more superiorly than $\text{DCDSR}_{RP}$ on all three datasets, thus demonstrating the effectiveness of the embedding collaborative perturbation.

 

\begin{table}[h]
\caption{Performance comparison between different embedding perturbation methods.}
\resizebox{\linewidth}{!}{
\begin{tabular}{cc|cc|c}
\hline Datasets & Metrics & $\text{DCDSR}_{RP}$ & $\text{DCDSR}_{CP}$ & $Improve. \%$ \\
\hline\hline \multirow{2}{*}{\begin{tabular}{c} 
Douban- \\
Book
\end{tabular}} & R@20 & 0.1725 & 0.1794
 & $+4.00 \%$ \\
 & $\mathrm{N} @ 20$ & 0.1664 & 0.1726  & $+3.73 \%$ \\
\hline \multirow{2}{*}{ Yelp } & R@20 & 0.1230 & 0.1346 & $+9.43 \%$ \\
 & $\mathrm{N} @ 20$ & 0.0746 & 0.0820 & $+9.92 \%$ \\
\hline \multirow{2}{*}{\begin{tabular}{c} 
Douban- \\
Movie
\end{tabular}} &  R@20 & 0.1053 & 0.1116
 & $+5.98 \%$ \\
 & $\mathrm{N} @ 20$ & 0.2553 & 0.2663 & $+4.31 \%$ \\
\hline
\end{tabular}}
    \label{CP VS RP}
\end{table}

\subsubsection{\textbf{AC-InfoNCE vs InfoNCE}} 
\begin{figure}
    \centering
    \includegraphics[width=\linewidth]{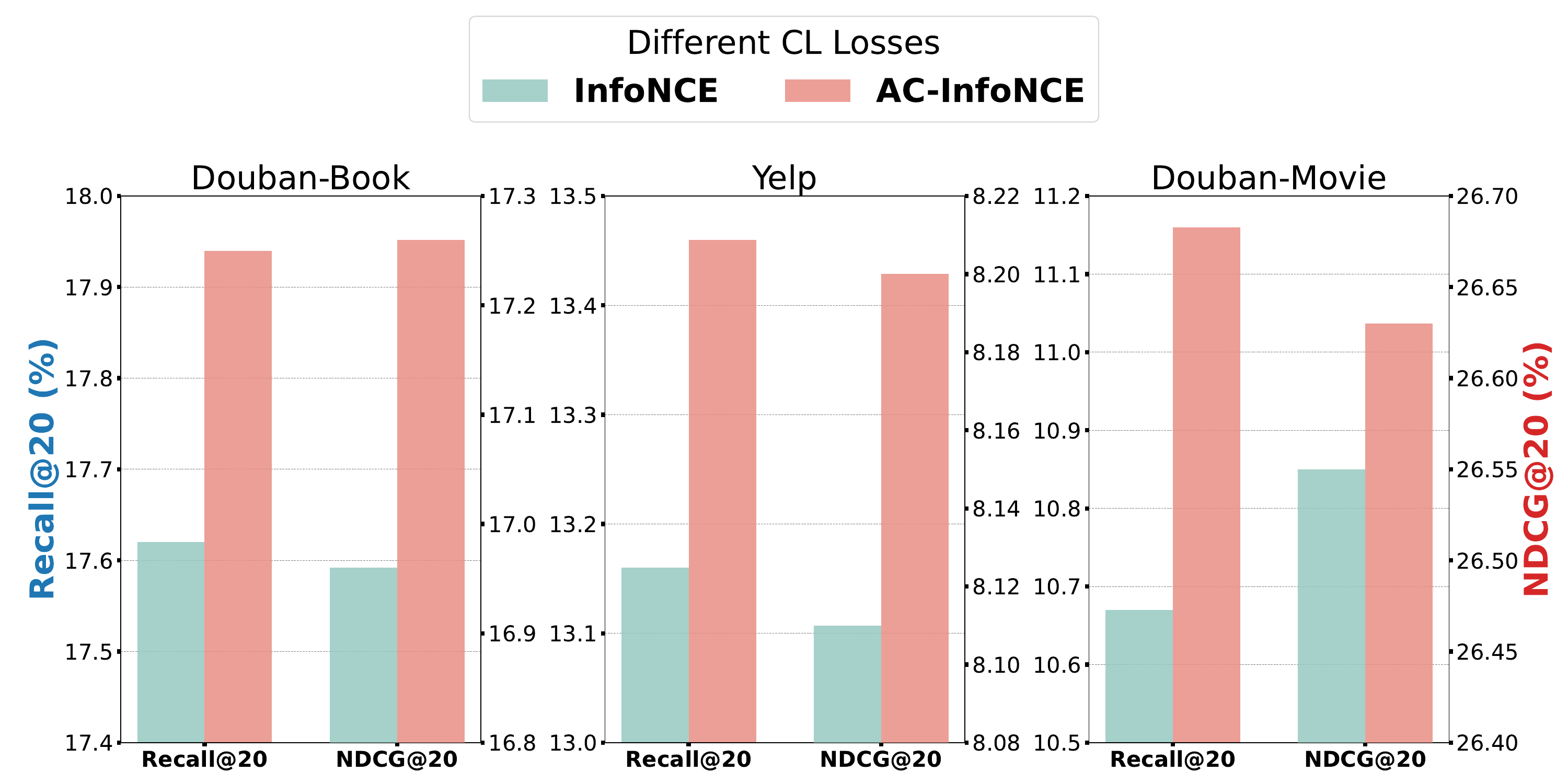}
    \caption{Performance comparison of DCDSR with InfoNCE and AC-InfoNCE.}
    \label{infonce_vs_ac_infonce}
\end{figure}

To verify the superiority of our proposed AC-InfoNCE, we compare the performance of DCDSR using AC-InfoNCE with that of DCDSR using InfoNCE. As shown in Fig. \ref{infonce_vs_ac_infonce}, on three datasets, replacing InfoNCE with AC-InfoNCE brings significant performance gains to DCDSR, confirming that AC-InfoNCE does allievate the gradient bias and data sparsity issues. The consistent improvement across different datasets demonstrates that AC-InfoNCE is a more powerful contrastive learning loss function and can be extended to other research fields. \\
\vspace{-1em}

\subsection{Noise Resistance Evaluation (RQ3)}
 To evaluate the noise resistance of DCDSR, we conduct noise attack experiments on DCDSR, DSL, SEPT and $S^2$-MHCN. We randomly generated fabricated user-item interactions with different ratios (i.e., 0.1, 0.2, and 0.3) to reconstruct the interaction data, and the variation of model performance with different noise ratios is shown in the following Fig. \ref{noise_ratio}. It can be seen that on both douban-book and yelp, the performance of DCDSR with different ratios of noise attack is less degraded than other models. 
 We attribute this to the following reasons: (1) The structure-level collaborative denoising module introduces denoising for interaction graph, which uses high-quality social information to play a supervised role on reconstructing a relatively clean interaction graph. This provides the model with a defensive capability in the face of interaction noise. (2) The embedding-space collaborative denoising module is based on contrastive learning, which allows interaction domain to effectively absorb the auxiliary information from the social domain, thus providing more self-supervised signals for the recommendation task, and enabling the model to have the ability to automatically identify noise. 

\subsection{Parameter Sensitivity Analysis (RQ4)} In this section, we explore the impact of some key hyperparameters on model performance. 
\subsubsection{Impacts of $\beta_s$ and $\beta_r$} 
\begin{figure}[t]
    \begin{minipage}[t]{0.33\linewidth}
        \centering
        \includegraphics[width=\textwidth]{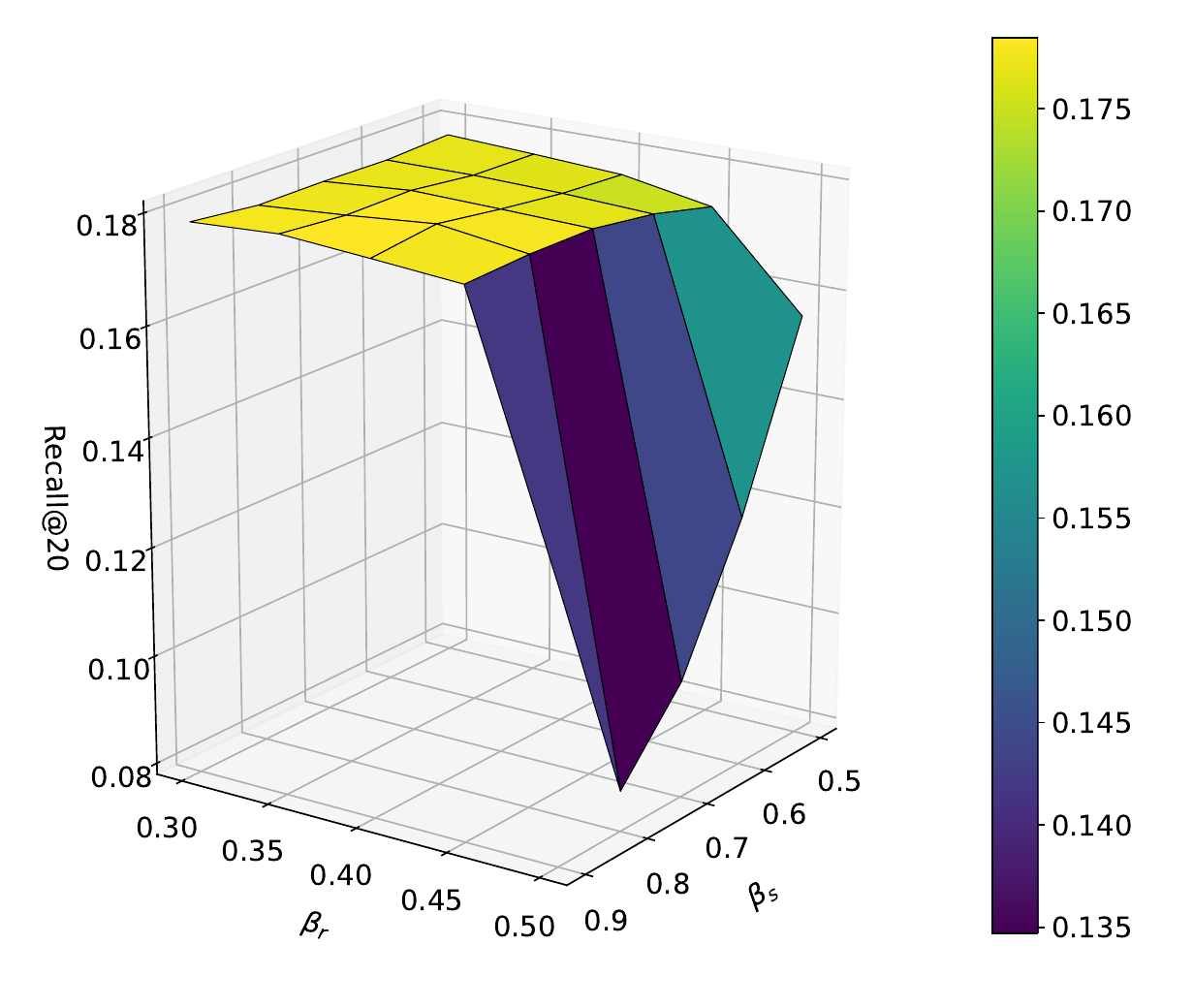}
        \centerline{(a) douban-book}
    \end{minipage}%
    \begin{minipage}[t]{0.33\linewidth}
        \centering
        \includegraphics[width=\textwidth]{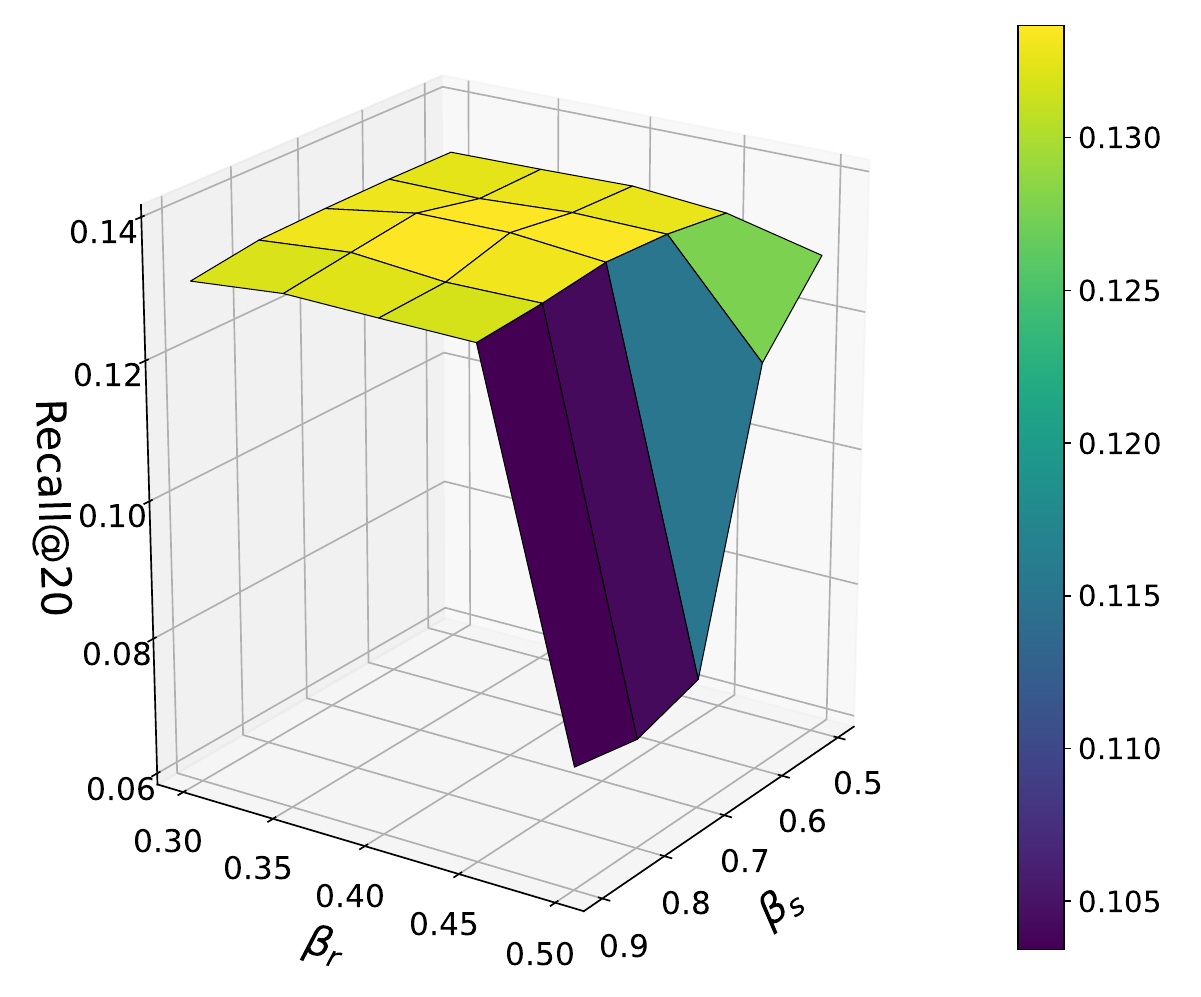}
        \centerline{(b) yelp}
    \end{minipage}%
    \begin{minipage}[t]{0.33\linewidth}
        \centering
        \includegraphics[width=\textwidth]{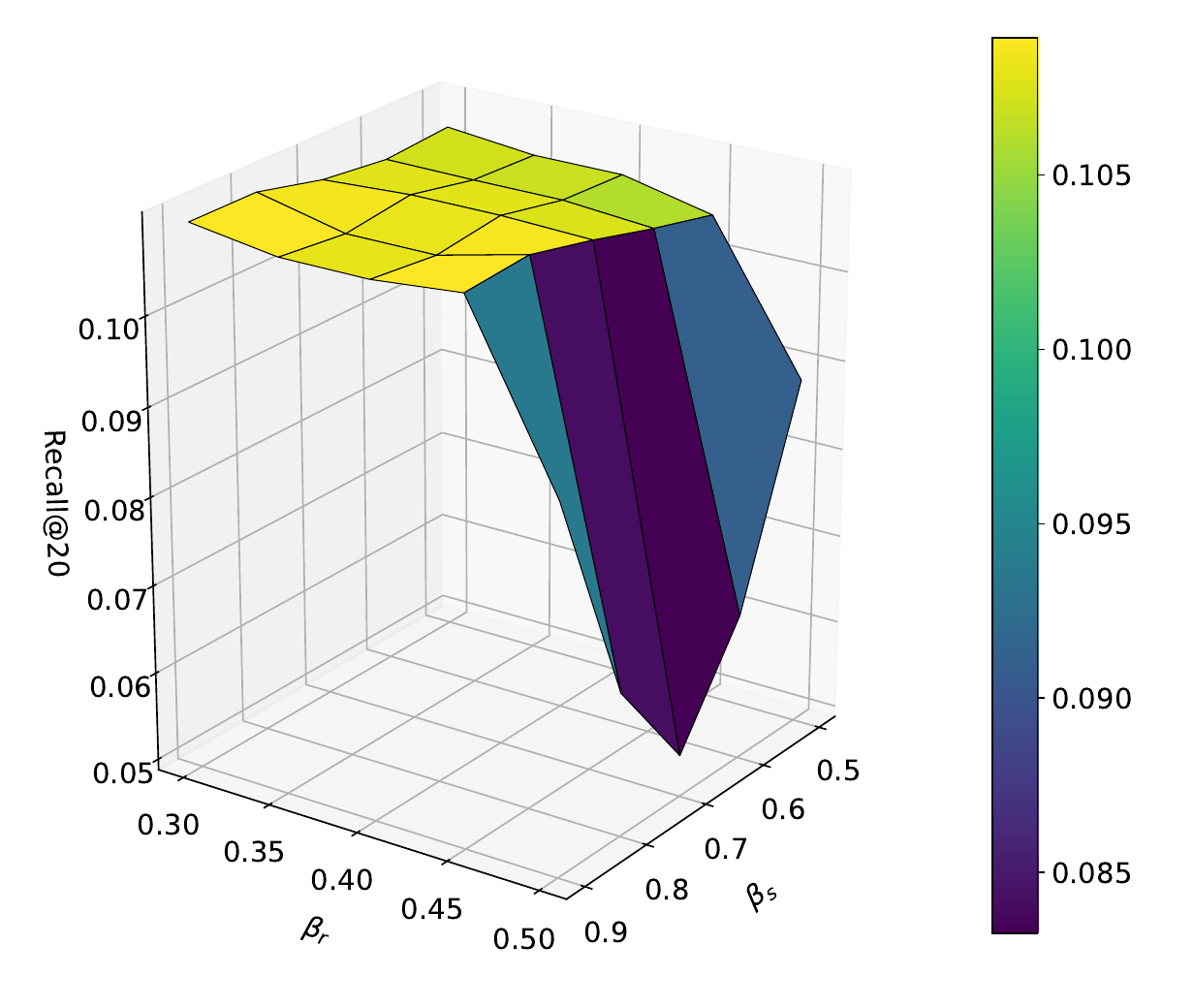}
        \centerline{(c) douban-movie}
    \end{minipage}
    \begin{minipage}[t]{0.33\linewidth}
        \centering
        \includegraphics[width=\textwidth]{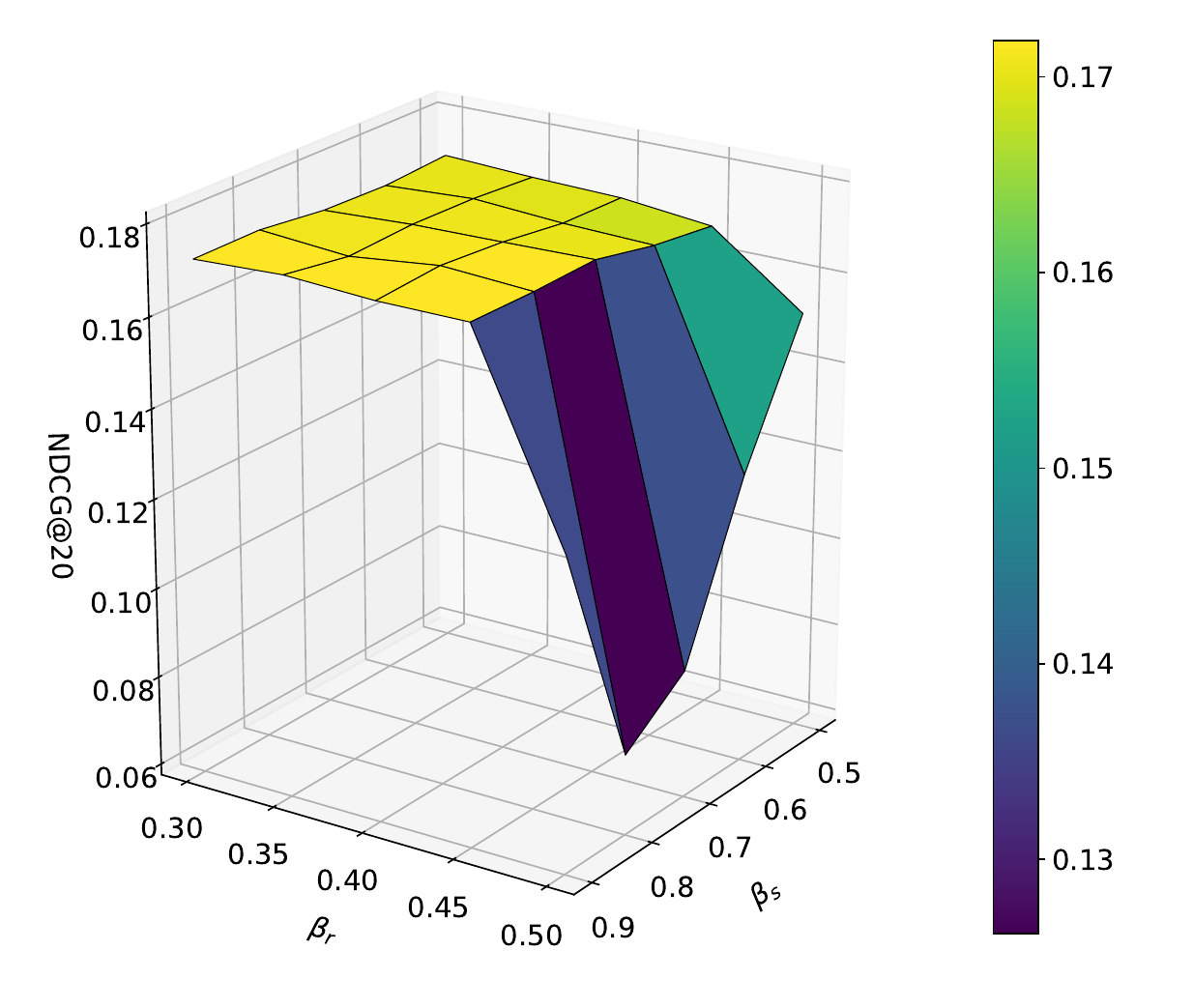}
        \centerline{(d) douban-book}
    \end{minipage}%
    \begin{minipage}[t]{0.33\linewidth}
        \centering
        \includegraphics[width=\textwidth]{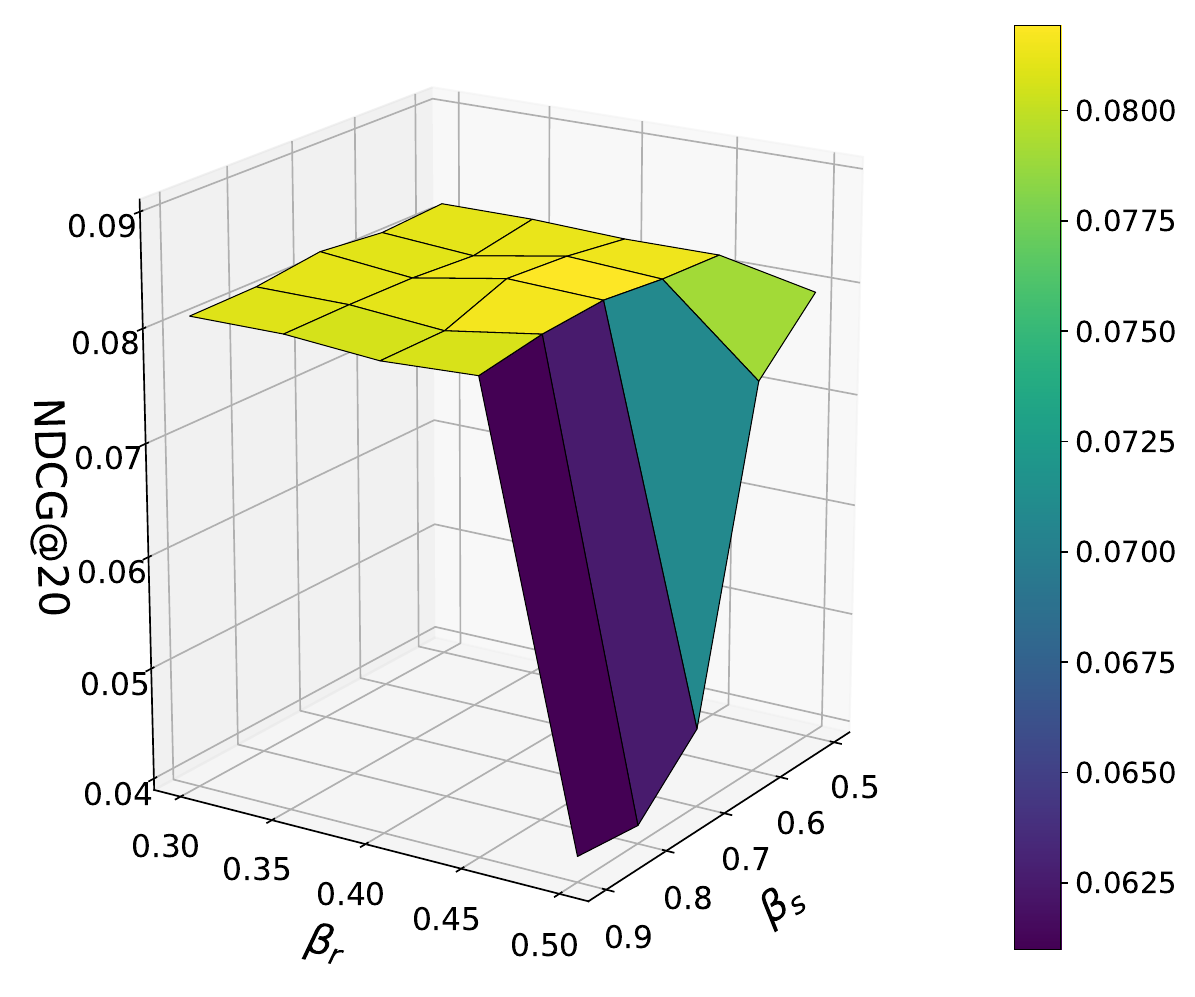}
        \centerline{(e) yelp}
    \end{minipage}%
    \begin{minipage}[t]{0.33\linewidth}
        \centering
        \includegraphics[width=\textwidth]{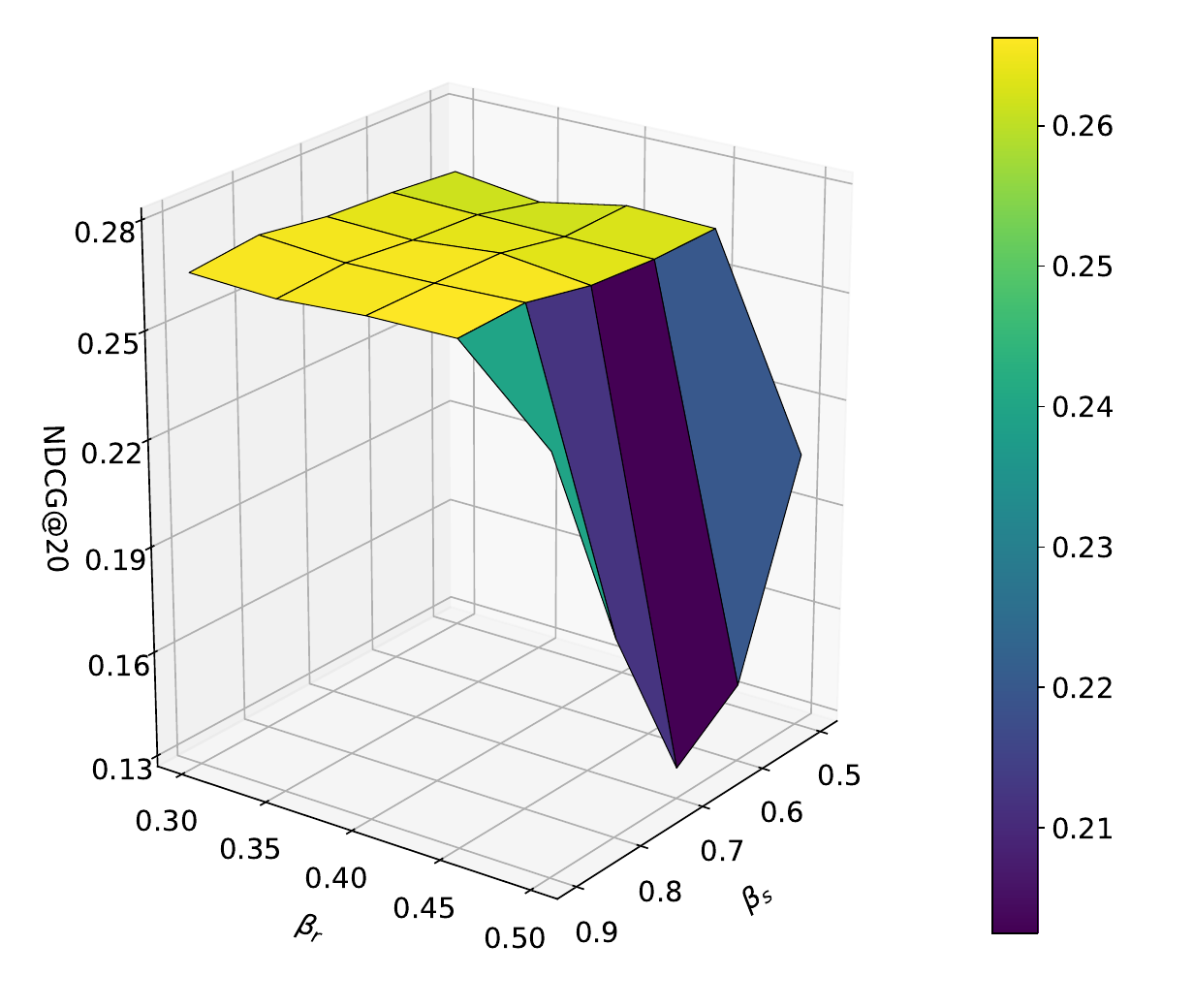}
        \centerline{(f) douban-movie}
    \end{minipage}
    \captionsetup{justification=raggedright}
    \caption{\textbf{Recommendation Recall@20 and NDCG@20 w.r.t. $\beta_s$ and $\beta_r$ on different datasets.}}
    \label{threshold}
\end{figure}

In the structure-level collaborative denoising module, the preference consistency threshold $\beta_s$ and the interaction compatibility threshold $\beta_r$ are crucial. They directly influence the retention or deletion of edges within the social network and interaction graph. To assess their impacts, we conduct experiments on three datasets, holding other parameters constant while varying $\beta_s$ and $\beta_r$ within the ranges [0.5, 0.6, 0.7, 0.8, 0.9] and [0.3, 0.35, 0.4, 0.45, 0.5], respectively. Fig. \ref{threshold} illustrates the effects of different $\beta_s$ and $\beta_r$ settings on Recall@20 and NDCG@20 metrics.

 \begin{figure*}[t]
    \begin{minipage}[t]{0.25\linewidth}
        \centering
        \includegraphics[width=\textwidth]{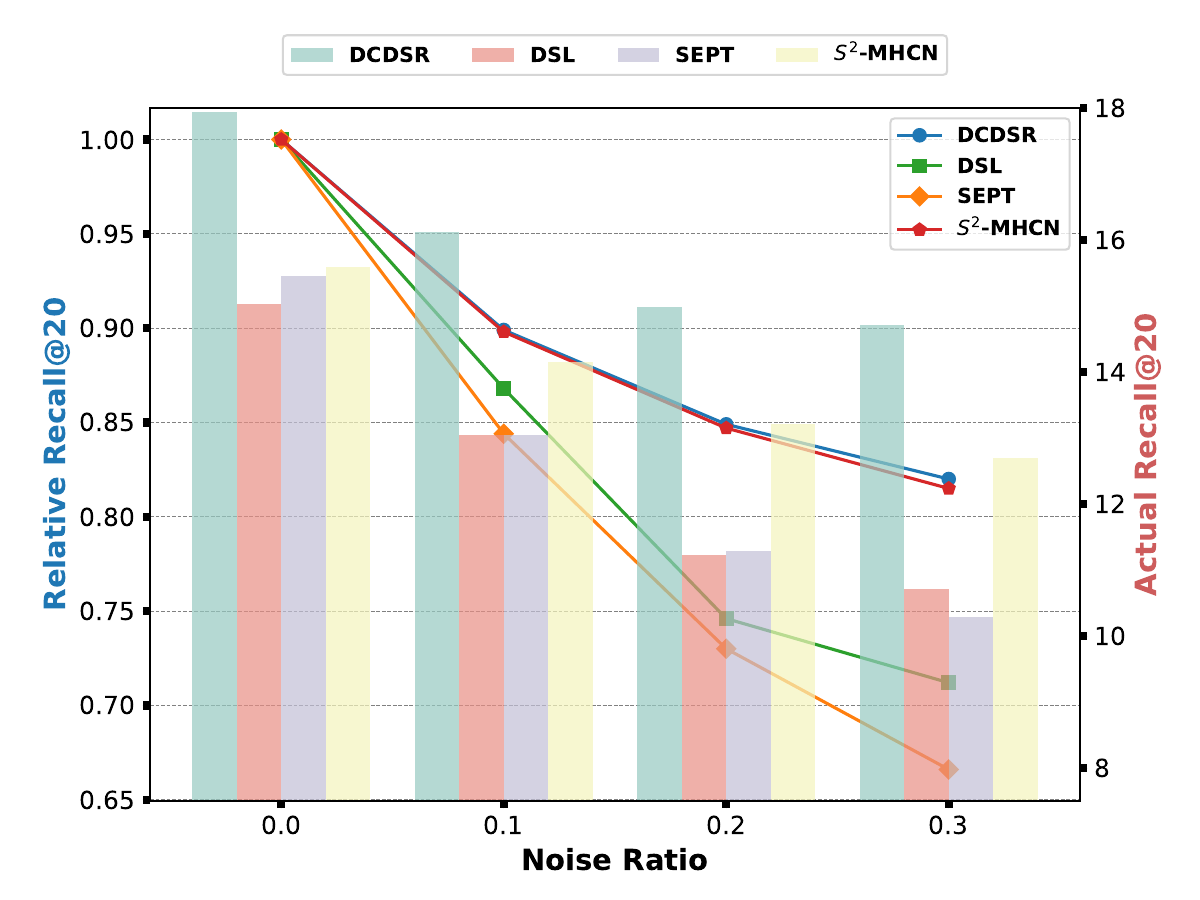}
        \centerline{(a) douban-book}
    \end{minipage}%
    \begin{minipage}[t]{0.25\linewidth}
        \centering
        \includegraphics[width=\textwidth]{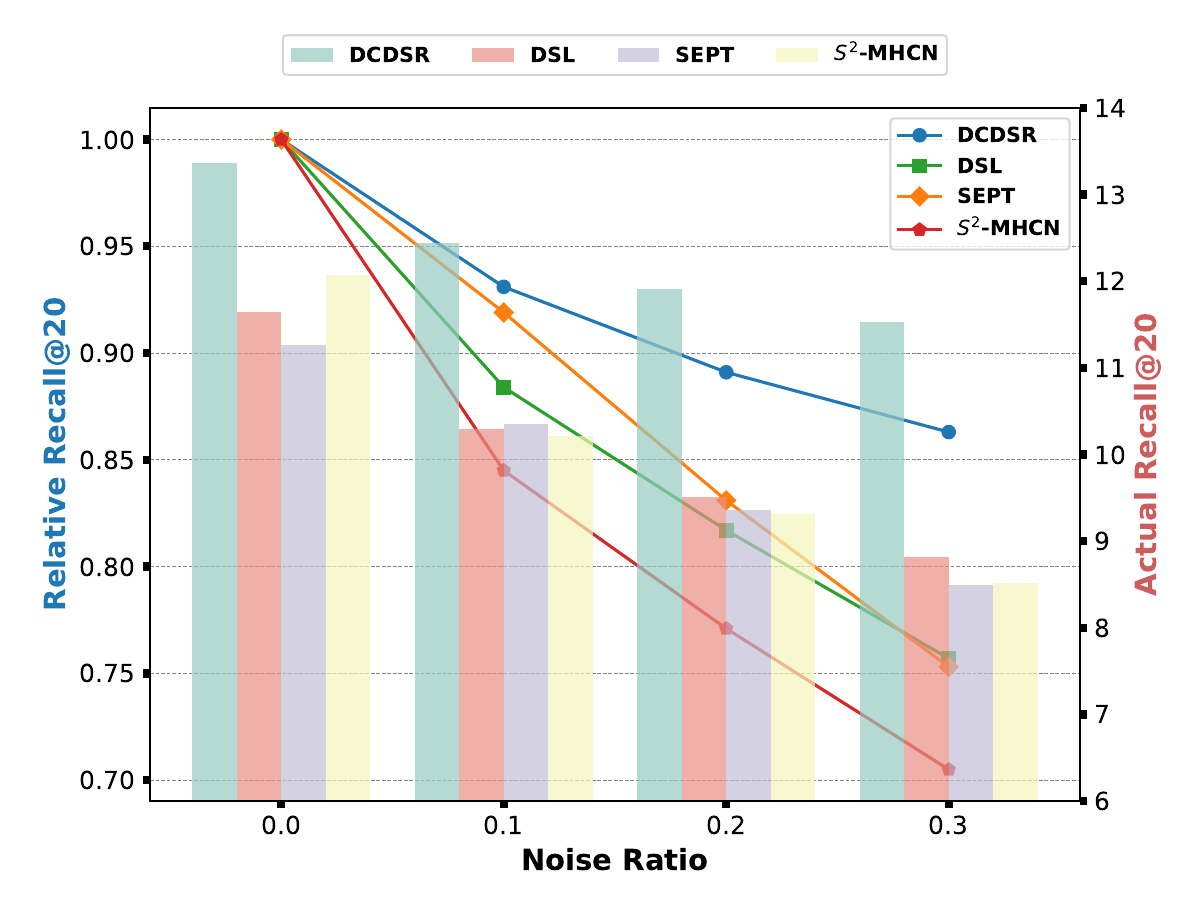}
        \centerline{(b) yelp}
    \end{minipage}%
    \begin{minipage}[t]{0.25\linewidth}
        \centering
        \includegraphics[width=\textwidth]{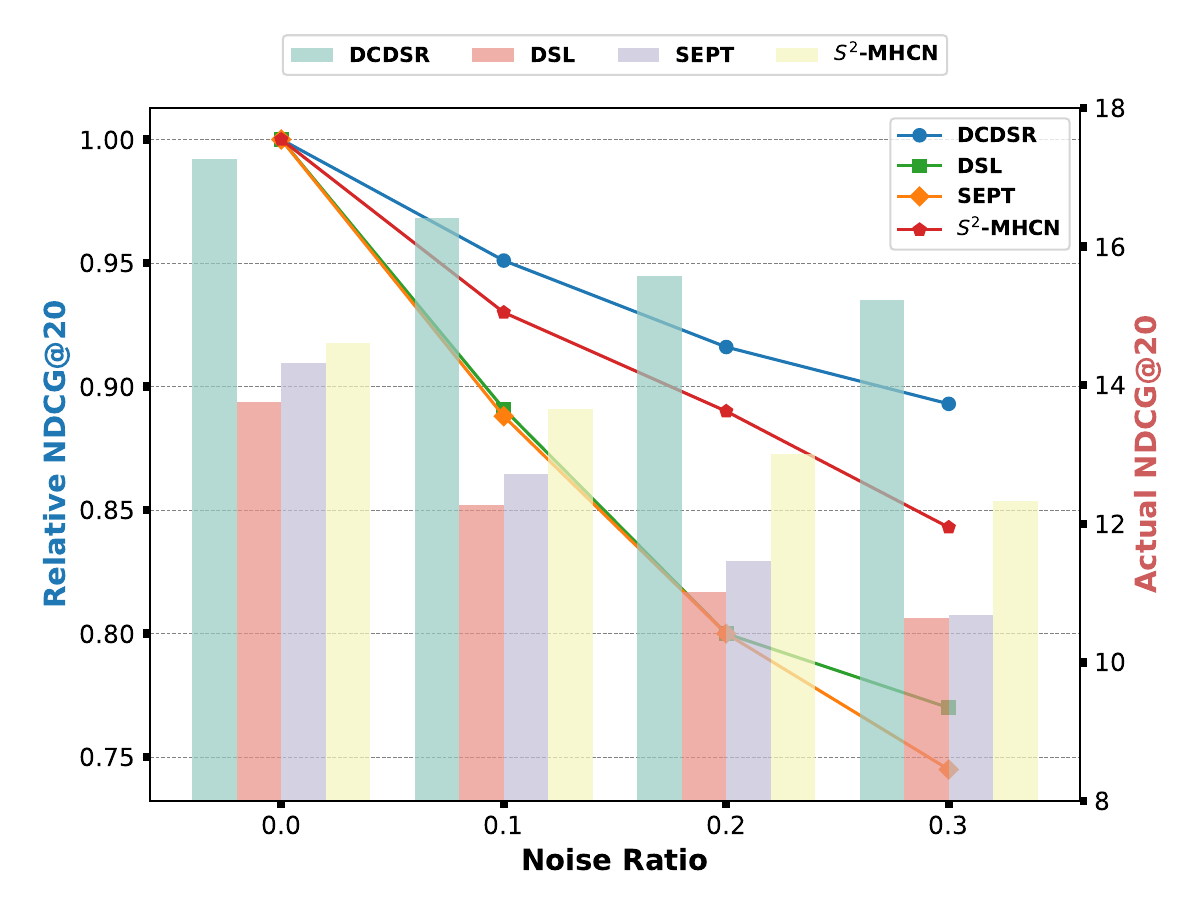}
        \centerline{(c) douban-book}
    \end{minipage}%
    \begin{minipage}[t]{0.25\linewidth}
        \centering
        \includegraphics[width=\textwidth]{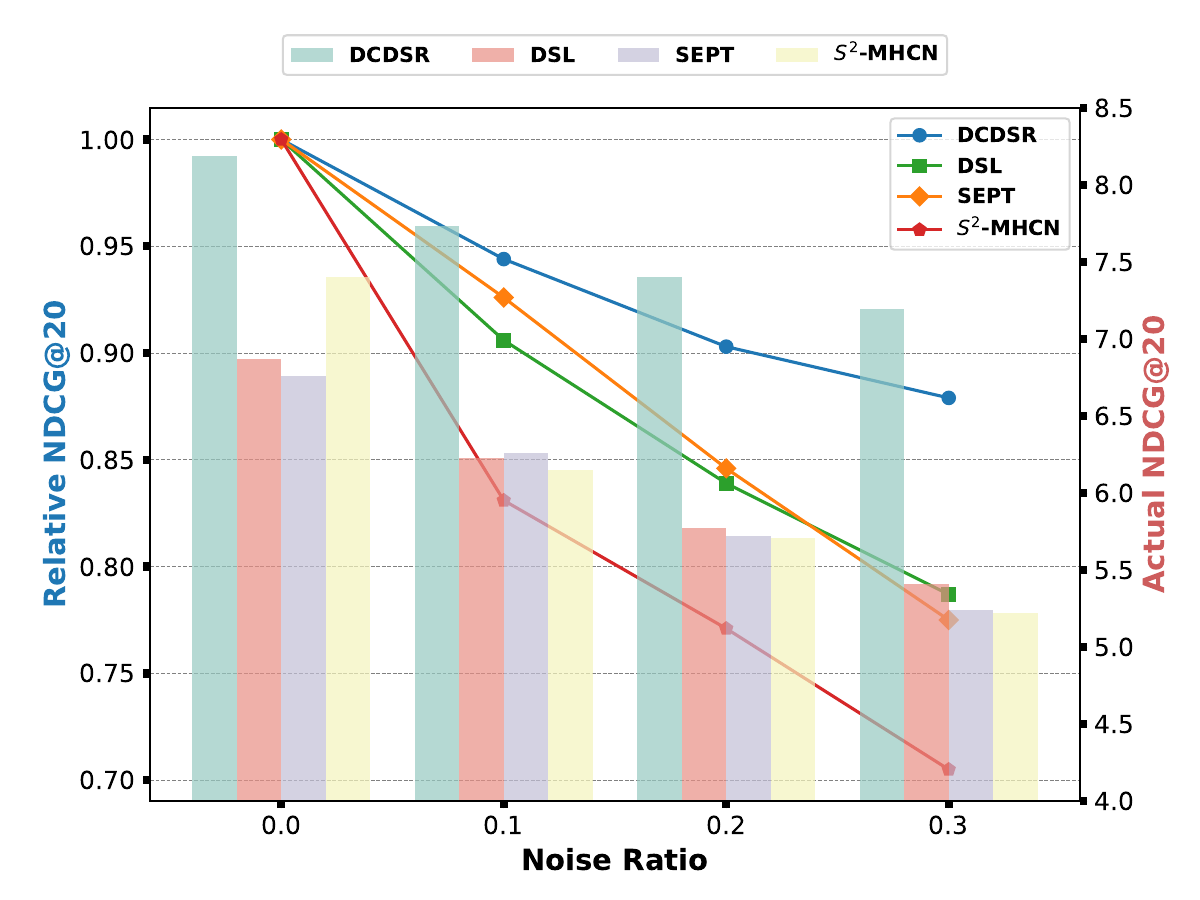}
        \centerline{(d) yelp}
    \end{minipage}
    \captionsetup{justification=raggedright}
    \caption{\textbf{Relative Recommendation Recall@20 and NDCG@20 w.r.t. different noise ratio on douban-book and yelp datasets.}}
    \label{noise_ratio}
\end{figure*}

Firstly, we analyze their impacts on model performance separately. For $\beta_s$, a clear trend is observed across all three datasets: as $\beta_s$ increases, DCDSR performance also improves, peaking at optimal values (0.8 for Douban-Book, 0.7 for Yelp, and 0.8 for Douban-Movie) before declining with further increases. This observation indicates that higher values of $\beta_s$, which maintain social relations with greater preference consistency, improve the quality of social network and thereby facilitate the denoising for interaction graph. Additionally, this leads to more reliable user social embeddings learned via GNN, which benefits collaborative denoising in the embedding space. The model performance associated with $\beta_r$ initially improves as the threshold increases, achieving an optimal value at 0.4 for Douban-Book, Yelp, and Douban-Movie. Beyond this peak, performance begins to decline. We believe that the deletion of edges for the original interaction graph needs to be careful, because a large $\beta_r$ will cause too many interaction edges to be deleted thus aggravating the data sparsity problem, while too small $\beta_r$ will result in noisy interaction data escaping filtered.

Subsequently, from a synergistic perspective, we conclude that DCDSR recursively repeats such pattern: ensuring that there is a large preference consistency (i.e., $\beta_s$) among social neighbors to get a social network that contains high-quality social information. Based on this, fusing social information to enhance the  user embeddings in interaction domain, and then removes interaction edges with low user-item compatibilty to refine interaction graph while ensuring that the supervisory signals are not drastically reduced.\\
\vspace{-1em}
\subsubsection{Impacts of $\lambda1$, $\lambda2$ and $\lambda_3$.} 
\begin{figure}[t]
    \begin{minipage}[t]{0.33\linewidth}
        \centering
        \includegraphics[width=\textwidth]{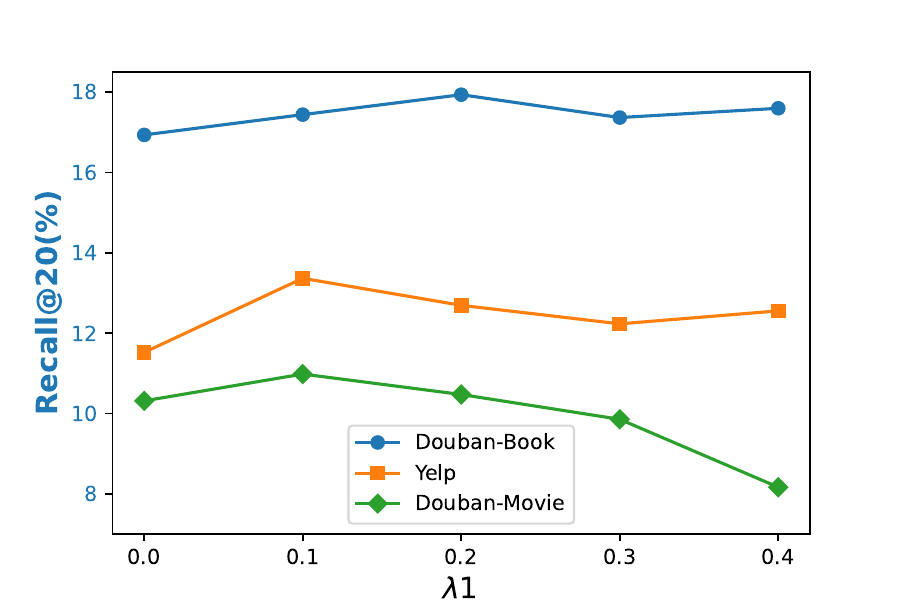}
        \centerline{(a) $\lambda1$}
    \end{minipage}%
    \begin{minipage}[t]{0.33\linewidth}
        \centering
        \includegraphics[width=\textwidth]{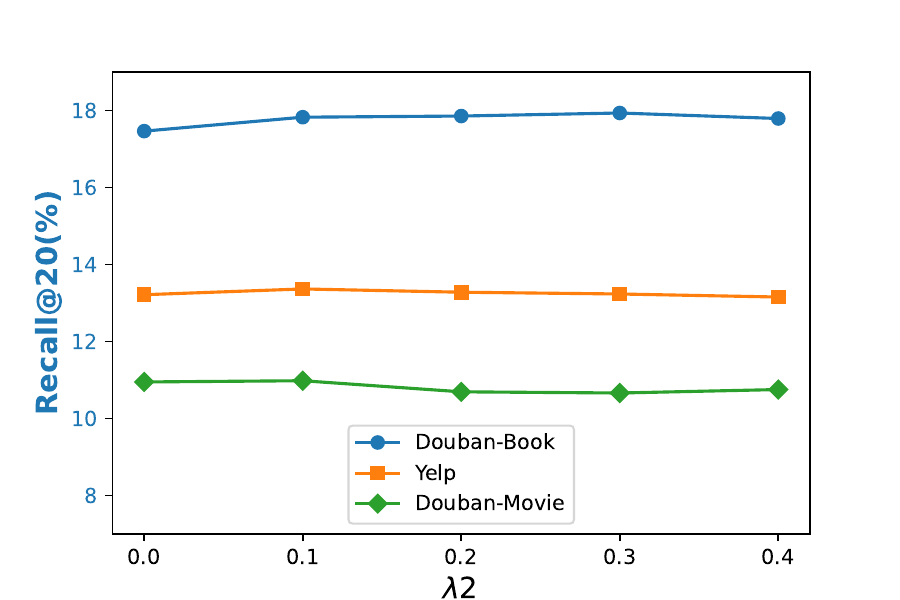}
        \centerline{(b) $\lambda2$}
    \end{minipage}%
    \begin{minipage}[t]{0.33\linewidth}
        \centering
        \includegraphics[width=\textwidth]{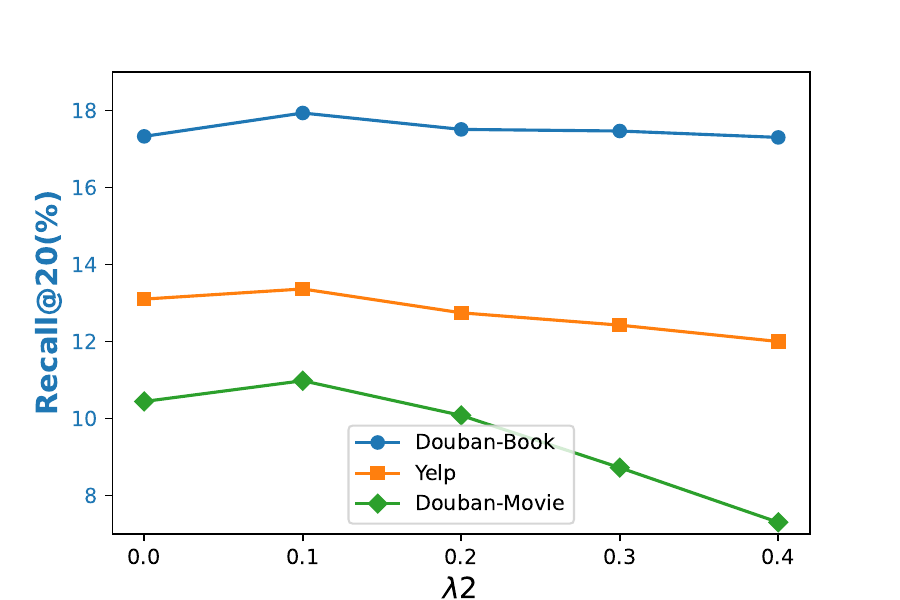}
        \centerline{(c) $\lambda3$}
    \end{minipage}
    \begin{minipage}[t]{0.33\linewidth}
        \centering
        \includegraphics[width=\textwidth]{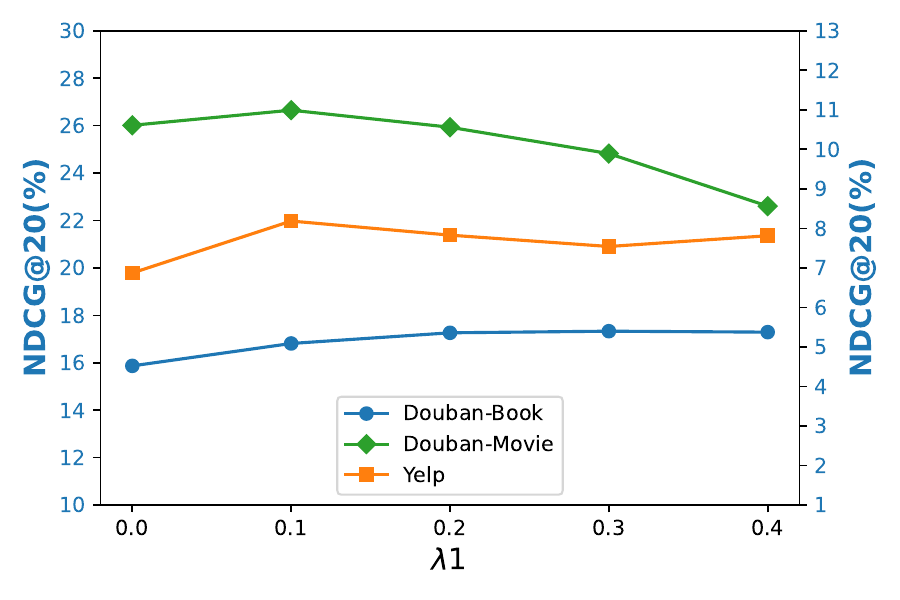}
        \centerline{(d) $\lambda1$}
    \end{minipage}%
    \begin{minipage}[t]{0.33\linewidth}
        \centering
        \includegraphics[width=\textwidth]{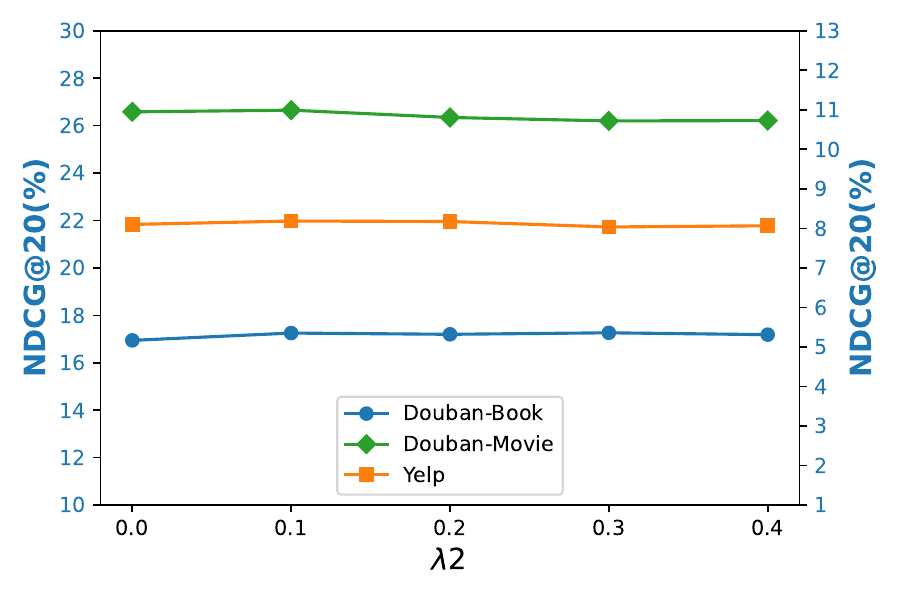}
        \centerline{(e) $\lambda2$}
    \end{minipage}%
    \begin{minipage}[t]{0.33\linewidth}
        \centering
        \includegraphics[width=\textwidth]{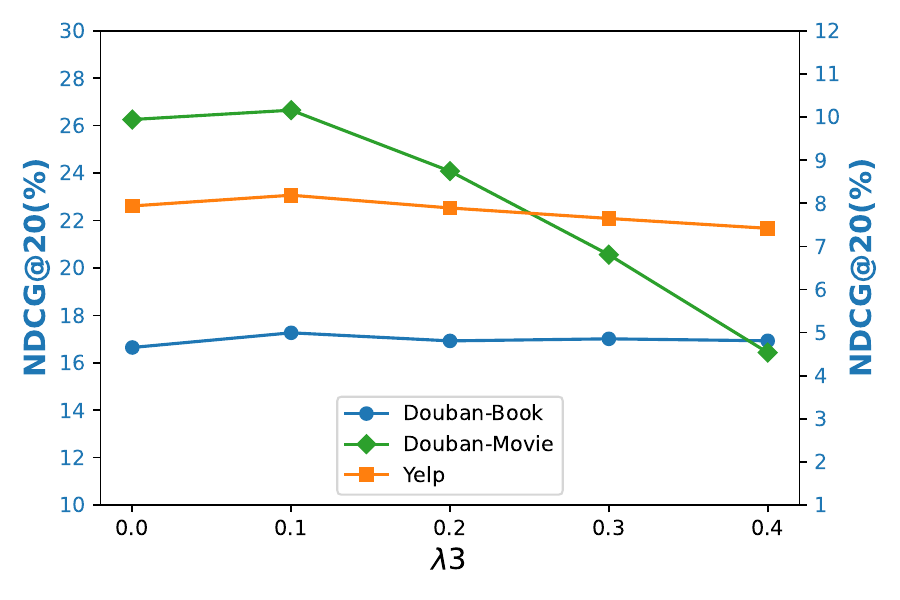}
        \centerline{(f) $\lambda3$}
    \end{minipage}
    \captionsetup{justification=raggedright}
    \caption{\textbf{Recommendation Recall@20 and NDCG@20 w.r.t. different weights of each contrastive loss on douban-book, yelp, and douban-movie datasets.}}
    \label{cl weight}
\end{figure}

In the embedding-space collaborative denoising module, three contrastive learning losses ($\mathcal{L}_{cl}^r$, $\mathcal{L}_{cl}^s$, $\mathcal{L}_{cl}^i$) guide the robust embedding learning process for the user/item. Assigning different weights to these three losses plays a crucial role in model training. Therefore, on all three datasets, we fixed the other parameters and adjusted them in the range of [0, 0.1, 0.2, 0.3, 0.4] to observe how the model performance changes. Observing Fig. \ref{cl weight}, on yelp and douban-movie, for three loss weights, the model obtains the best performance when they are all taken to 0.1. On douban-book, the model achieves its best performance when $\lambda1$, $\lambda2$, and $\lambda_3$ are taken to 0.2, 0.3, and 0.1. This suggests that douban-book dataset may contain more noisy interactions and social relations and thus requires a higher degree of denoising. Overall, for each weight, on all three datasets, the model performance increases firstly and then decreases as it increases, and the model performs poorly when it takes zero. This suggests that each contrastive learning loss is effective for embedding space denoising, while it should not be over-represented to prevent it from inhibiting the optimization of recommendation task.\\


\section{Conclusion and Future Work}
In this paper, we explore the noise issues in social recommendation and propose a novel denoising social recommendation model, called DCDSR. DCDSR contains two denoising modules, namely, structure-level collaborative denoising and embedding-space collaborative denoising. The structure-level collaborative denoising process is applied to construct cleaner social network and interaction graph with a collaborative denoising paradigm. The embedding-space collaborative denoising module simulates the raw embeddings unaffected by noise cross-domain diffusion with dual-domain embedding collaborative perturbation and learn denoised embeddings with contrastive learning. Ultimately, DCDSR learns high-quality and robust user/item embeddings for recommendation through multi-task joint training. Experimental results on various datasets demonstrate that DCDSR outperforms the state-of-the-art models including graph-based recommendation, graph-based social recommendation and denoising social recommendation, validating the superiority of DCDSR in dealing with the noise issues. More importantly, DCDSR not only solves the noise issue in social network, but also collaboratively reduces the noise of interaction data, which truly plays the auxiliary role of social network for collaborative filtering.

In the future, we will conduct a more granular investigation into the noise issues in social recommendation to better perform denoising task in social recommendation. Furthermore, both noise identification methods and noise filtering techniques are worthwhile directions for research.

\section*{Acknowledgments}
This work is supported by the National Science Foundation of China (No. 62272001 and No. 62206002), and Hefei Key Common Technology Project (GJ2022GX15).

%

\bibliographystyle{IEEEtran}

\bibliography{ref}

\newpage

 




\vfill

\end{document}